\documentclass[12pt]{article}
\usepackage{amsmath,amssymb,graphicx}

\makeatletter
\renewcommand\section{\@startsection {section}{1}{\z@}%
                                 {-3.5ex \@plus -1ex \@minus -.2ex}
                                   {2.3ex \@plus.2ex}%
                                   {\normalfont\large\bfseries}}
\renewcommand\subsection{\@startsection{subsection}{2}{\z@}%
                                   {-3.25ex\@plus -1ex \@minus -.2ex}%
                                     {1.5ex \@plus .2ex}%
                                     {\normalfont\bfseries}}
\renewcommand\subsubsection{\@startsection{subsubsection}{3}{\z@}%
                                   {-3.25ex\@plus -1ex \@minus -.2ex}%
                                     {1.5ex \@plus .2ex}%
                                     {\normalfont\itshape}}
\makeatother

\newcommand{\Letter}{
    \setlength{\textwidth}{7in}
    \setlength{\textheight}{9.5in}
    \hoffset=-0.75in
    \voffset=-1.15in }

\Letter

\newcommand{\mpl}{M_{\rm Pl}}
\newcommand{\spec}{{\cal P}}
\newcommand{\fin}{{\rm e}}
\newcommand{\fnl}{f_{\rm NL}}
\newcommand{\tip}{{\rm end}}

\def\cK{{\mathcal{K}}}

\def\cG{{\mathcal{G}}}

\def\cO{{\mathcal{O}}}

\def\bz{\bar{z}}

\def\al{\alpha}
\def\ga{\gamma}

\def\nn{\nonumber}

\def\r2{{\sqrt{2}}}
\def\h{{\eta}}

\def\h0{\hat{h}}
\def\Vr0{\hat{V}_{r}}
\def\Vp0{\hat{V}_{\phi}}

\def\r2{\sqrt{2}}

\def\nn{\nonumber}

\def\ss{\sigma_\star}
\def\phim{\phi_\mu}
\def\sepy{|\mathbf{y}-\mathbf{\bar{y}}|}
\def\mbV{{\mathbb{V}}}
\def\phir{{\phi_r}}
\def\phit{{\phi_\tau}}

\begin{document}

\begin{titlepage}
\setcounter{page}{1} \baselineskip=15.5pt \thispagestyle{empty}
\begin{flushright}
\parbox[t]{2in}{
MAD-TH-08-10 \\
CERN-PH-TH/2008-146}
\end{flushright}

\vfil


\begin{center}
{\Large \bf {Systematics of Multi-field Effects \\
\vskip 8pt
at the End of Warped Brane Inflation}}
\end{center}
\bigskip

\begin{center}
{Heng-Yu Chen$^{1}$, Jinn-Ouk Gong$^{1}$, and Gary Shiu$^{1,2}$}
\end{center}

\begin{center}
    \textit{$^1$Department of Physics,
     University of Wisconsin-Madison,
     Madison, WI 53706-1390, USA}
     \vskip 3pt
     \textit{$^2$PH-TH Division, CERN, CH-1211 Geneva 23, Switzerland}
\end{center}
\bigskip \bigskip \bigskip \bigskip

\begin{center}
{\bf Abstract}
\end{center}

We investigate in the context of brane inflation the possibility of additional light
scalar fields generating significant power spectrum and non-Gaussianities at the end
of inflation affecting the CMB scale observations. We consider the specific
mechanism outlined by Lyth and describe the necessary criteria for it to be
potentially important in a warped throat. We also discuss different mechanisms for
uplifting the vacuum energy which can lead to different dominant
contributions of the
inflaton potential near the end of inflation. We then apply such
criteria to one of the most detailed brane inflation models to date, and show that inflation
can persist towards the tip of the throat, however for the specific stable
inflationary trajectory, the light residual isometry direction becomes degenerate.
We also estimate the effects for other inflationary trajectories with non-degenerate
residual isometries.

\vfil
\begin{flushleft}
\today
\end{flushleft}

\end{titlepage}

\newpage



\tableofcontents

\newpage

\section{Introduction}
\paragraph{}

Inflation~\cite{inflation} has emerged as the standard paradigm describing physics
of the very early universe. Besides addressing several fine-tuning issues in big
bang cosmology such as the flatness and horizon problems, it provides a framework to
explain the origin of structure and the cosmic microwave background (CMB)
anisotropy~\cite{books}. While there is a
plethora of effective field theory based models of inflation~\cite{Lyth:1998xn},
many outstanding questions in inflationary cosmology {\it require} a fundamental
microscopic description. Conversely,
recent observations of the CMB and large scale structure~\cite{observation}
lead us to increasingly precise measurements of the inflationary parameters.
These measurements provide us with an
exciting window to probe physics at ultra-high energies~\cite{reconstruction},
far higher than what
current and upcoming terrestrial accelerators can reach. Thus inflationary
cosmology has become the perfect arena for fundamental theory to meet experiment.

String theory is currently our leading candidate for a quantum theory of gravity.
Thus it is worthwhile to explore explicit realizations of inflation within this
framework. In this paper, we will focus on one of the most well developed
inflationary scenarios in string theory, i.e. $D$ brane inflation~\cite{DvaliTye}
(see also Refs.~\cite{Dvali:2001fw,Burgess:2001fx}; for reviews, see Ref.~\cite{reviews} and
references therein), where the inflaton field is identified with the position of a
space-filling mobile $D$ brane, usually a $D3$ brane, in a warped six dimensional
manifold~\cite{KKLMMT}. In the original scenario of
Refs.~\cite{DvaliTye,Dvali:2001fw,Burgess:2001fx,reviews,KKLMMT}, an additional
$\overline{D3}$ was introduced to drive inflation. The $\overline{D3}$ brane is
localized by the RR fluxes at the tip of a warped throat, thus inflation  proceeds
as the mobile $D3$ is attracted by a weak $D3$-$\overline{D3}$ Coulombic force to
move slowly along the warped direction. However, it was also noted in
Ref.~\cite{KKLMMT} that because the volume modulus of the compactification couples
non-trivially to the canonical inflaton, its stabilized value gives additional
Hubble scale correction to the inflaton mass, causing the well known $\eta$
problem~\cite{Copeland:1994vg}.

An important step towards addressing the $\eta$ problem explicitly in this concrete
setting was recently made in Refs.~\cite{Baumann0, Baumann1} (see also
Refs.~\cite{Panda:2007ie, McGill1, Krause:2007jk, Pajer:2008uy}). The key ingredient in
the construction was the one loop
threshold correction to the non-perturbative superpotential obtained in
Ref.~\cite{Gaugino} (see also Refs.~\cite{Ganor,BHK}). In Ref.~\cite{KKLT} and other
stabilized compactifications, non-perturbative effects are often introduced to
stabilize moduli. In the context of Ref.~\cite{KKLT}, such effects come from
instantons on a stack of $D7$ branes (or Euclidean $D3$ branes). Interestingly, the
non-perturbative moduli stabilizing force also turns out to give the dominant
contribution to the inflaton potential\footnote{Although the Coulombic force is
subdominant in comparison to the moduli stabilizing force, a $\overline{D3}$ brane
was still introduced to end inflation.}. This contribution arises because the mobile
$D3$ brane backreacts on the moduli stabilizing $D7$ branes. The correction depends
on the holomorphic four cycles within the conifold on which the $D7$ branes wrap.
The embedding of the $D7$ branes breaks the isometry of the deformed conifold, and
thus the inflationary trajectory depends sensitively on the choice of the embedding
function {defining the loci of the $D7$ branes}. As a result, explicit slow-roll
models have been constructed by a ``delicate'' tuning of the microscopic
compactification parameters.

While the broken angular isometry directions are stabilized by the coordinate
dependent non-perturbative superpotential, for a given $D7$ brane embedding, there
are typically residual isometries preserved by the resultant scalar potential. The
potential for the fields associated with these isometries remain flat during the
inflationary epoch and so they can take arbitrary values without affecting the
inflationary trajectory. Being almost massless, their quantum fluctuations give rise to a nearly
scale invariant isocurvature perturbation spectrum. As argued by Lyth and
collaborators in Refs.~\cite{Lyth,Lyth1}, these isocurvature perturbations can be
converted to the curvature perturbations at the end of inflation. In the context of
$D$ brane inflation, inflation ends when the open string tachyon condenses between
the mobile $D3$ and $\overline{D3}$. The critical value of the canonical inflaton at
which inflation ends $\phi_\tip$ depends on the residual symmetries as they enter
into the tachyon potential. Since $\phi_\tip$ picks up spatial dependence through
the quantum fluctuations of the light residual symmetries, inflation can end on a
spatial slice of non-uniform energy density. {As we will see, this is the case for
instance when the inflaton potential is dominated by the moduli stabilizing force
towards the end of inflation. Thus,} one could in principle expect potentially
significant contribution to the power spectrum and non-Gaussianities at the end of
inflation.

In this paper, we study these multi-field effects at the end of brane inflation, and
outline the necessary conditions for them to be significant. We then perform a case
study for the setup considered in Ref.~\cite{Baumann1}, by explicitly calculating
the canonical inflaton potential near the tip of the deformed conifold, and
demonstrate that inflation can persist in this region provided that the
$D3$-$\overline{D3}$ Coulombic attraction becomes subdominant. We also discuss
various mechanisms to uplift the vacuum energy which results in a subdominant
Coulombic potential all the way to the tip of a warped throat. We also show that the
angular stable inflationary trajectory for the specific $D7$ brane embedding
\cite{Kuper} used in Ref.~\cite{Baumann1} can be extended to the entire deformed
conifold. However, along the specific trajectory considered in Ref.~\cite{Baumann1},
we will see explicitly that the corresponding residual angular isometries have
vanishing proper separations at the tip. Thus, for this specific $D7$ embedding, {no
significant contribution to the curvature perturbation is generated at the end of
inflation}. This implies that while multi-field effects can in principle be
significant in brane inflation, they can only happen with other $D7$ embeddings, or
with more than one stacks of $D7$ branes present.

This paper is organized as follows. In Section~\ref{sec_D3brane}, we review the
basic setup of flux compactification and brane inflation, in order to set up our
notation. Readers who are familiar with the above topics can skip this section. In
Section~\ref{Lytheffect}, we recast the mechanism proposed in Ref.~\cite{Lyth} in
the context of brane inflation in a warped throat, and outline the necessary
conditions for it to take place. In Section~\ref{sec_scenario}, we discuss various
possible uplifting mechanisms in a warp throat and propose a natural scenario for an
uplifted potential to realize the effect of Ref.~\cite{Lyth}. In
Section~\ref{sec_endofinf}, we explicitly calculate the canonical inflaton potential
near the tip of the deformed conifold and the resulting slow-roll parameters. The
degeneracy of the residual isometries  will also be shown. We end with some
discussions in Section~\ref{Discussion}. We relegate most of the calculational
details in a number of Appendices.

\section{$D3$ brane in warped compactifications}
\label{sec_D3brane}
\setcounter{equation}{0}
\paragraph{}

We will consider  warped compactification of type IIB string theory in four
dimensions~\cite{GKP} (see also earlier works on warped IIB vacua
\cite{Dasgupta:1999ss,Verlinde:1999fy,Greene:2000gh}), with the following metric ansatz
\begin{equation}
ds^2 = e^{2A(y)} e^{-6 u(x)} g_{\mu\nu} dx^\mu dx^\nu + e^{-2 A(y)} e^{2 u(x)}
\tilde{g}_{mn} dy^m dy^n \, ,
\label{eq:GKPmetric}
\end{equation}
where $e^{A(y)}$ is the warp factor sourced by branes and fluxes, and $e^{u(x)}$ is
the Weyl rescaling factor required to decouple the overall volume modulus from the
four dimensional graviton, which can be taken as $\cO(1)$. The internal metric
$\tilde{g}_{mn}$ is taken to be that of a compact six dimensional Calabi-Yau space.
In addition to the ansatz (\ref{eq:GKPmetric}), we choose the bulk RR and NS-NS
fluxes of type IIB supergravity to respect four dimensional  Lorentz
invariance (and self-duality in the case of the five form flux),
\begin{align}\label{flux_G3}
G_3 &\equiv F_3 - \tau H_3 = \frac{1}{6} G_{mnp}\ dy^m\wedge dy^n\wedge dy^p \, ,
\\
\widetilde{F}_5 & = (1+*)d\alpha(y) \wedge \sqrt{|g_4|} dx^0\wedge dx^1\wedge dx^2
\wedge dx^3 \, ,
\end{align}
where we combined the NS-NS and RR three forms $H_3$ and $F_3$ with the complex
axio-dilaton $\tau\equiv C_0+ie^{-\phi}$ into the complex combination $G_3$.

We are interested in the background BPS solutions of the equations of motion which
impose the following relations on the fluxes~\cite{GKP},
\begin{align}
\alpha(y) & = e^{4A(y)} \, ,
\\
*_6 G_3 & = i G_3 \, ,
\end{align}
such that the complexified three form flux is imaginary self-dual.

\subsection{Four dimensional effective theory}
\label{4DEffTh}
\paragraph{}

At energy scales much lower than the Kaluza-Klein mass scale, the effective theory
for this warped background is described by four dimensional $\mathcal N = 1$
supergravity. The scalar fields of our theory consist of closed string moduli,
including the complex structure moduli, axio-dilaton and K\"ahler moduli, as well as
open string moduli, such as the positions of $D3$ branes and $D7$ branes. The
flux-induced superpotential~\cite{GVW}
\begin{equation}
W_\mathrm{GVW} = \int G_3\wedge \Omega
\label{eq:GVW}
\end{equation}
stabilizes the complex structure moduli and axio-dilaton as described in
Ref.~\cite{GKP}, where $\Omega$ is the holomorphic $(3,0)$ form of the unwarped
Calabi-Yau space. Lifting (\ref{eq:GVW}) to F-theory, we see that bulk and $D7$
worldvolume fluxes can also stabilize the positions of $D7$ branes as well.  We will
assume for the rest of the paper that these moduli are stabilized by (\ref{eq:GVW})
and its F-theory lift, and we will work at energies below the scale of this
stabilization. The stabilized complex moduli give rise to a constant contribution to
the superpotential,
\begin{equation}\label{VGVW0}
W_0 \equiv \left(\int G_3 \wedge \Omega\right)_0 \, .
\end{equation}

The remaining closed and open string moduli consist of the K\"ahler moduli,
associated with the sizes of holomorphic four cycles, and the positions of $D3$
branes in the internal space.  For simplicity we will consider a single K\"ahler
modulus $\rho = \sigma + i \varsigma$, and denote the location of the $D3$ brane in
the compact space by three complex coordinates $z^\alpha$ with $\alpha = 1,2,3$.  In
the presence of a $D3$ brane, the K\"ahler potential for the $D3$ brane fields and
the K\"ahler modulus is~\cite{DG}
\begin{equation}\label{eq:KahlerPotential}
\kappa^2 \cK(\rho,z^\alpha, \bz^{\alpha}) = -3 \log \left[ \rho+\bar{\rho}-\gamma k
\left( z^\al,\bar{z}^\al \right) \right] \equiv -3 \log U(r,\sigma) \, ,
\end{equation}
where
\begin{align}
\gamma = & \frac{\sigma_0T_3}{3M_P^2} \, ,
\\
\kappa^2 = & \mpl^{-2} = 8\pi G \, ,
\end{align}
$k(z^\al,\bar{z}^\al)$ is the geometric K\"ahler potential for the metric on the
Calabi-Yau, and $\sigma_0$ is the stabilized value of $\sigma$ when the $D3$ brane
is at its stabilized configuration: see Ref.~\cite{D3vacua} for more details. It is
important to note that there are many subtle issues involved in the derivation of
the low energy effective action for warped compactifications. These issues discussed
in, e.g. Refs.~\cite{Giddings:2005ff,Frey:2006wv,Burgess:2006mn} raise some
concerns about the validity of the above conjectured warped Kahler potential
\cite{DG}  in the strong warping limit, though some recent progress has been made
towards this end \cite{Shiu:2008ry,Douglas:2008jx}.

In type IIB compactifications the flux superpotential (\ref{eq:GVW}) does not depend
on the K\"ahler moduli, so we need other ingredients to stabilize these fields.  One
mechanism for stabilizing the K\"ahler moduli is to include non-perturbative effects
through gaugino condensation on a stack of $D7$ branes or a Euclidean $D3$ brane
instanton. Branes wrapping a four cycle associated with a K\"ahler modulus $\rho$
produce a non-perturbative contribution to the superpotential which depends on
$\rho$ and the $D3$ brane position $z^\alpha$ of the form
\begin{equation}\label{DefWnp}
W_\mathrm{np} = A(z^\al) e^{-a \rho} \, ,
\end{equation}
with $a = 2\pi/n$, where $n>1$ for gaugino condensation on $D7$ branes and $n=1$ for
a Euclidean $D3$ brane. The prefactor $A(z^\al)$ is a holomorphic function and can
be written as~\cite{Gaugino, Ganor, BHK}
\begin{equation}\label{DefAz}
A(z^\al) = A_0 \left[ \frac{f(z^\al)}{f(0)} \right]^{1/n} \, ,
\end{equation}
where $A_0$ depends on the stabilized complex structure moduli and has mass
dimension $3$. The dependence on the position of $D3$ branes shows up through the
embedding function $f(z^\al) = 0$ of the four cycle in the Calabi-Yau space, where
$f(0)$ represents the value of the embedding function when the $D3$ brane is
stabilized.

The total superpotential
\begin{equation}
\label{eq:SuperW}
W =  W_0 + A_0 \left[ \frac{f(z^\al)}{f(0)} \right]^{1/n} e^{-a \rho} \, ,
\end{equation}
and the K\"ahler potential (\ref{eq:KahlerPotential}) give rise to the $F$-term
contribution to the scalar potential which depends on the K\"ahler moduli and the
$D3$ positions,
\begin{equation}
V_{F}(\sigma,z^\al,\bz^\al)=e^{\kappa^2\cK} \left[
\cK^{\Sigma\Omega}D_{\Sigma}W\overline{D_{\Omega}W} - 3\kappa^2|W|^2 \right] \, .
\label{DefVF}
\end{equation}
Substituting the general superpotential (\ref{eq:SuperW}) as well as the explicit
expression for the inverse metric $\cK^{\Sigma\Omega}$ solved in Ref.~\cite{McGill1}
into (\ref{DefVF}), the explicit form for $V_F(\sigma,z^\al)$ is given by
\begin{align}
V_F(\sigma,z^\al,\bz^\al) = & \frac{\kappa^2}{3[U(r,\sigma)]^2} \left\{ \left[ U(r,\sigma) +
\gamma k^{\gamma\bar\delta} k_{\ga} k_{\bar{\delta}} \right] |W_{,\rho}|^2 - 3
\left( \overline{W}W_{,\rho} + W\overline{W}_{,\bar{\rho}} \right) \right\}
\nonumber\\
& + \frac{\kappa^2}{3[U(r,\sigma)]^2} \left[ \left(
k^{\al\bar\delta}k_{\bar\delta}\overline{W}_{,\bar{\rho}}W_{,\al} +
k^{\bar{\al}\delta}k_{\delta}W_{,\rho}\overline{W}_{,\bar{\al}} \right) +
\frac{1}{\gamma} k^{\al\bar\beta}W_{,\al}\overline{W}_{,\bar{\beta}} \right] \, ,
\label{explicitVF}
\end{align}
where a subscript of a letter with a comma denotes a partial differentiation with
respect to the corresponding component. Clearly the scalar potential depends on the
detailed form of the little K\"ahler potential $k(z^\al,\bar{z}^\al)$ and its
derivatives, as well as the holomorphic $D7$ brane embedding function $f(z^\al)$.

\subsection{Warped deformed conifold}
\paragraph{}

The localized fluxes and sources can backreact on the geometry and generate a
non-trivial warp factor $e^{A(y)}$~\cite{GKP}
\begin{equation}\label{eq:warpfactor}
\widetilde{\nabla}^2 e^{4A} = e^{2A} \frac{|G_3|^2}{12{\rm Im}\tau} + 2 e^{-6A}
(\partial e^{4A})^2 + \frac{\kappa_{10}^2}{2} e^{2A} \left( T^m{}_m - T^\mu{}_\mu
\right)^\mathrm{local} \, ,
\end{equation}
where $T^m{}_n$ is the stress energy tensor of `localized' sources such as $D3$ and
$D7$ branes. When fluxes are turned on along the $A$ and $B$ cycles in the
neighborhood of a conifold point in the internal space,
\begin{align}
\frac{1}{2\pi\alpha'}\int_A F_3 = & 2\pi M \, ,\label{quantM}
\\
\frac{1}{2\pi\alpha'}\int_B H_3 = & -2\pi K \, ,\label{quantK}
\end{align}
they generate a strongly warped `throat'. The complex structure modulus $\epsilon^2
= \int_A \Omega$ of the conifold is stabilized at an exponentially small
value~\cite{GKP}
\begin{equation}\label{defparameter}
\epsilon^2=\sqrt{2}\upsilon_0^{3/4}(g_s M\alpha')^{3/2}a_0^3 \, ,
\end{equation}
with
\begin{align}
\upsilon_0 \approx & 0.718050 \, ,
\\
a_0 = & \exp\left(-\frac{2\pi K}{3 g_s M}\right) \, .
\end{align}
The geometry is that of a {\it warped deformed conifold}, whose construction in
supergravity is known as the Klebanov-Strassler (KS) throat~\cite{KS}.
Notice that in our definition of the deformation parameter $\epsilon^2$, the exponential
warp factor $a_0$ explicitly appears.
The six
dimensional deformed conifold can be described by a deformation of the embedding of
the singular conifold in ${\mathbb C}^4$ as
\begin{equation}\label{Defepsilon}
\sum_{A=1}^4 (z^A)^2 = \epsilon^2 \, ,
\end{equation}
where we will use the $SO(4)$ rotational symmetry of the
coordinates ${z^\al}$ to make the deformation parameter $\epsilon$ real.

The detailed metric of the warped deformed conifold is given in Appendix
\ref{app_conifold}, so here we just note that far from the `tip'
of the throat where the deformation is concentrated, the metric is simply that of a
singular conifold,
\begin{equation}
ds_6^2 \approx \frac{3}{2} \left( dr^2 + r^2 d\Omega_{T^{1,1}}^2 \right) =
d\hat{r}^2 + \hat{r}^2 d\Omega_{T^{1,1}}^2 \, ,
\label{Defsingconmetric}
\end{equation}
where the space $T^{1,1}$ is a Einstein-Sasaki metric with the topology of
$S^2\times S^3$ and we define $\hat{r}^2 = 3r^2/2$ for notational simplicity. Near
the tip of the throat, $S^2$ shrinks to zero size while $S^3$ remains finite with
its size given by the deformation parameter with the metric
\begin{equation}
ds_6^2 \approx \epsilon^{4/3} \left( d\tau^2 + \tau^2 d\Omega_{2} + d\Omega_{3}
\right) \, .
\label{eq:TipMetric}
\end{equation}
Here the parameter $\tau\in \mathbb R$ is related to the radial coordinate $r$ and
the embedding coordinates $z^\al$ via
\begin{equation}\label{Defr}
\epsilon^2 \cosh\tau = \sum_{A=1}^4 |z^A|^2 = r^3 \, .
\end{equation}
We hope our readers do not confuse $\tau$ here with the IIB {axio-dilaton shown in
(\ref{flux_G3}). Throughout the remaining text, $\tau$ denotes this coordinate.}
The expressions for the complex embedding coordinates $z^A$ given in terms of real
coordinates are listed in Appendix \ref{app_conifold}.

\subsection{$D3$ brane dynamics}
\label{D3Dyn}
\paragraph{}

We are interested in the dynamics of mobile $D3$ branes in the background discussed
above. For slowly moving $D3$ branes, the kinetic term is derived from the pull-back
of the bulk deformed conifold metric
\begin{equation}
S_{3} = -\frac{1}{2} T_{3} \int d^4x \sqrt{|g_4|}\ e^{-4u} g^{\mu\nu} \partial_\mu
Y^\alpha \partial_\nu \overline{Y}^{\bar{\beta}} \tilde{g}_{\alpha\bar{\beta}} \, ,
\label{eq:kineticterms}
\end{equation}
where $\tilde{g}_{\alpha\bar{\beta}} = \partial_\alpha \partial_{\bar{\beta}} k$
denotes the bulk deformed conifold metric. In general (\ref{eq:kineticterms}) is a
non-linear sigma model, so it is not always straightforward to canonically normalize
all the fields simultaneously into the form
\begin{equation}
S_\mathrm{norm} = \int d^4x \sqrt{|g_4|} g^{\mu \nu} \partial_\mu \phi^\alpha
\partial_\nu \bar{\phi}^{\bar{\beta}}\delta_{\alpha\bar{\beta}}\, .
\end{equation}
In particular, far from the tip of the throat where we can write the metric as
(\ref{Defsingconmetric}), we can write the kinetic term for a spatially homogenous
$D3$ brane as
\begin{equation}\label{D3conifoldaction}
S_{3} = \frac{3}{2}T_{3} \int d^4x \sqrt{|g_4|}\ e^{-4u} \left[ \dot{r}^2 + r^2
d\dot{\Omega}_{T^{1,1}}\right] \, ,
\end{equation}
where a dot indicates a derivative with respect to the time coordinate $t$. For
motion only in the large radial direction, we can identify
\begin{equation}
\phi_r(t) \equiv \sqrt{\frac{3}{2}T_3} e^{-2u} r(t) = \sqrt{T_{3}}\ e^{-2u}
\hat{r}(t)
\label{Inflatonr}
\end{equation}
as the canonically normalized scalar field in the radial direction far from the tip.
Similarly using (\ref{eq:TipMetric}), near the tip of the throat the internal metric
is of the form given by (\ref{eq:TipMetric}) and the kinetic term becomes
\begin{equation}\label{ActionD3}
S_{D3} = T_{3} \int d^4x \sqrt{|g_4|}\ e^{-4u} \epsilon^{4/3}\left( \dot{\tau}^2 +
\tau^2 d\dot{\Omega}_{2} + d\dot{\Omega}_{3} \right) \, .
\end{equation}
Again, for the motion only in the small radial ($\tau$) direction, we can identify
\begin{equation}
\phi_\tau(t) \equiv \sqrt{\frac{T_3}{2}} \epsilon^{2/3} e^{-2u}\ \tau(t)
\label{Inflatontau}
\end{equation}
as the canonically normalized scalar field near the tip. Note that we have focused
on two regions in the deformed conifold, where the canonical inflaton can be defined
as a simple function of local coordinates. However, in general, the definition of
the canonical inflaton valid for the entire deformed conifold can be more involved,
and it should interpolate between the two asymptotic limits (\ref{Inflatonr}) and
(\ref{Inflatontau}).
Furthermore, we restricted
our analysis above to cases where the multiple field trajectory is composed of a
single field (the radial direction) and consider only quantum fluctuations in the
light angular directions.
More generally, however, the inflationary system consists of multiple fields for
which simple analytic expressions of the canonical inflaton fields in terms of the
coordinates is not possible.

In the setup of Refs.~\cite{KKLMMT,Baumann1}, inflation proceeds as a mobile $D3$
brane is driven towards the tip of the warped deformed conifold, where a
$\overline{D3}$ brane is located. The $D3$-$\overline{D3}$ interactions are through
two different potentials. In the closed string channel, $D3$ and $\overline{D3}$
interact gravitationally via the potential
\begin{equation}\label{DefVD3barD3}
V_{D3\overline{D3}}(|\mathbf{y}-\mathbf{\bar{y}}|) =
\frac{D(|\mathbf{y}-\mathbf{\bar{y}}|)}{[U(r,\sigma)]^2} \, ,
\end{equation}
where
\begin{equation}\label{DyD3barD3}
D(|\mathbf{y}-\mathbf{\bar{y}}|) = D_0 \left( 1 -
\frac{3D_0}{16\pi^2T_3^2|\mathbf{y}-\mathbf{\bar{y}}|^4} \right) \, .
\end{equation}
Here $D_0=2T_3a_0^{4}$ is the warp factor at the
tip of the warped deformed conifold. One should remember that
$|\mathbf{y}-\mathbf{\bar{y}}|$
contains {\em both} radial and angular separations\footnote{The potential $V_{D3\overline{D3}}$
as written diverges when $\sepy\to 0$. However as demonstrated in Ref.~\cite{Interbrane},
the Coulombic potential gets smoothed out to finite value through regularization as the
separation becomes local string length.}.
Furthermore, in the open string channel, which becomes relevant as the
$D3$-$\overline{D3}$ separation approaches the local string length, tachyon condensation
develops, whose contribution to the overall scalar potential can be derived from open string
one-loop computation is given by
\begin{equation}
V_\mathrm{tach}(|\mathbf{y}-\mathbf{\bar{y}}|) = T_3 |T|^2 \left(
|\mathbf{y}-\mathbf{\bar{y}}|^2 - a_0^2\alpha '\right) + \cdots \, ,
\label{Tachyonpotential}
\end{equation}
where $T$ is the complex tachyon field.
The dot ellipsis indicates that the tachyon potential can receive higher order
contributions
in $D3$-$\overline{D3}$ separation $\sepy$~\cite{Interbrane}. %
%
While the high order terms in the $D$ brane separation can change the behavior of
the tachyon potential within the tachyon condensation surface, there is hardly an
$e$-folds at such small separation that such higher order contributions can be
ignored.
To estimate the range the tachyon condensation
surface occupies in the coordinate space, we can consider near the tip, where the
local geometry approaches $\mathbb{R}\times S^2\times S^3$. The $D3$-$\overline{D3}$
separation then becomes
\begin{equation}\label{smallsep}
|\mathbf{y}-\mathbf{\bar{y}}|^2\approx \epsilon^{4/3} \left[\tau^2 +
\tau^2(\Delta\Omega_2)+ \Delta\Omega_3\right] \, .
\end{equation}
Here $\tau$ is related to $r$ via (\ref{Defr}) and in this coordinate
$\overline{D3}$ radial position is $\tau=0$, and $\Delta \Omega_2$ and
$\Delta\Omega_3$ denote the finite angular separations between $D3$ and
$\overline{D3}$ on $S^2$ and $S^3$, respectively.
One should also note that in
addition to the $\overline{D3}$ at the tip of the deformed conifold, there can be additional
distant $\overline{D3}$ or other supersymmetry breaking sources, e.g.
$D7$ with supersymmetry breaking worldvolume flux, present in the bulk. Their
presence also increases the potential energy and needs to be taken into account: in
fact they will play an important role in our subsequent discussion.

In the presence of $D3$ or $D7$ branes which wrap on a specific supersymmetric four
cycle in the throat and generate non-perturbative superpotential, some of the
angular coordinates which correspond to broken isometries are stabilized by the
$F$-term scalar potential $V_F$. Furthermore the stabilized values of these
directions are in fact the same for $D3$ and $\overline{D3}$ \cite{D3vacua,
NSUSYD3vacua}. However there can also be residual isometry direction(s) which remain
light compared with the canonical inflaton. Thus generally inflation ends when these
fields reach the tachyon condensation surface given by
\begin{equation}
\epsilon^{4/3} \left[\tau^2+\tau^2(\Delta\Omega_2^{\rm res})+\Delta\Omega_3^{\rm
res}\right]=a_0^2\alpha' \, .
\label{tachyonsurface}
\end{equation}
Here $\Delta \Omega_2^{\rm res}$ and $\Delta \Omega_3^{\rm res}$ indicate that the
only varying angular coordinates correspond to the residual isometry directions, and
their precise expressions depend on the specific embeddings.

As the deformed conifold is usually attached to a compact bulk Calabi-Yau manifold,
which contains additional ISD fluxes that further break these flat residual
isometries, these can possibly give masses to the corresponding $D3$ and
$\overline{D3}$ fields. To analyse such effects for $D3$, we can consider a probe
$D3$ and use gauge/string duality \cite{Bulkeffects1} (building on earlier
works~\cite{Ceresole:1999zs,Ceresole:1999ht}), the symmetry breaking can be encoded
by deforming the probe worldvolume theory with irrelevant operators. However the
consistent equations of motion would then require such terms to be vanishing, that
is for $D3$, the bulk flux {\textit {does not}} generate masses for the residual
symmetry fields. For $\overline{D3}$, such bulk flux generates perturbation in its
action through the dependence on the warp factor, an estimate for such effect was
given in Ref.~\cite{Bulkeffects2}: this generates a mass to the residual isometry
fields for $\overline{D3}$ of the order
\begin{equation}
m_\mathrm{bulk}^2 \sim \frac{a_0^{n}}{g_sM\alpha'} \, ,
\label{bulkmass}
\end{equation}
with $n \ge 3.29$, so that the bulk mass for $\overline{D3}$ residual isometry is
exponentially suppressed, and we still have an approximate
isometry\footnote{Attentive readers may also note that there can be further
contribution to the potential for the $D3$ residual isometry direction coming from
(\ref{DefVD3barD3}), which can give an effective mass of
$\mathcal{O}\left(a_0^2\right)$. However in the scenario which will be described
later, such term will be decoupled.}.

\section{The residual isometries and the Lyth effect}
\label{Lytheffect}
\setcounter{equation}{0}
\paragraph{}

In this section, we will give a general discussion on the mechanism proposed 
in Ref.~\cite{Lyth}, which can potentially generate significant contributions to the
curvature perturbation at the end of inflation due to the presence of the light
residual isometry fields. Furthermore we will also outline the necessary criteria
for such effect to take place in a warped throat.
 Some earlier related discussions in the context of brane inflation
appeared in Refs.~\cite{Lyth1,Leblond:2006cc}, though as we will see,
 our results differ in details.
In the following, we will refer to this additional contribution to the inflationary
perturbation at the end of inflation as the Lyth effect.

To begin our discussion, let us first estimate the maximum
value at which the residual isometry direction(s) can reach on the tachyon surface. For
simplicity we consider the situation where only a single residual isometry $\Theta$
is present\footnote{Here we use $\Theta$ to highlight such a special residual
angular direction and in general it should be a function of the usual angular
coordinates given in Appendix \ref{app_conifold}, whose specific form is dictated by specific
$D7$ brane embedding.}. The tachyon condensation surface is given by
(\ref{tachyonsurface}), from which we can estimate the maximum $D3$-$\overline{D3}$
angular separation $\Delta\Theta_\mathrm{\rm c}$ in the residual isometry direction
for the inflationary trajectory to reach the tachyon surface in field space. This
occurs when $\tau = 0$ and $D3$ reaches non-vanishing $S^3$ at the tip of deformed
conifold. Simple algebra then gives
\begin{equation}\label{Maxtheta}
\Delta\Theta_\mathrm{\rm c} = \frac{1}{\Gamma_3\sqrt{g_s M}} \, .
\end{equation}
Here we have used the definition $1/T_3 = (2\pi)^3g_s{\alpha'}^2$ and $\Gamma_3$
denotes a measure factor on $S^3$ which depends on the angular stable trajectory for
the specific $D7$ embedding. We have also absorbed the $\cO(1)$ numerical factor in the
definition of $\epsilon^2$ into $\Gamma_3$.
It is worth noting that the maximum angular separation $\Delta \Theta_c$ is not warp
factor suppressed because of the $a_0^2$ factor in $\epsilon^{4/3}$.
This is in contrast to the singular conifold case where $\Delta \Theta_c$ is
suppressed by $a_0$, in which case
the angular range is exponentially small.
While the factor ${1}/\sqrt{g_sM}$ is generally
small, the measure factor $1/\Gamma_3$ can be large, and whether $\Delta\Theta_{\rm
c}$ constitutes a fine-tuned initial condition needs to be examined on a case-by-case
basis.
If $\Delta\Theta_{\rm c}$ exceeds the allowed value for $\Delta \Theta$,
tachyon condensation would necessarily take place away from $S^3$ at $\tau
> 0$\footnote{A small mass due to bulk fluxes on the $\overline{D3}$ residual isometry
direction discussed earlier may change the story. However it is proportional to
$a_0^{3.29}$, which is even smaller than the possible effective mass of
$\cO\left(a_0^2\right)$ coming from the Coulombic term. Therefore if we can ignore
the Coulombic contribution in our proposed scenario discussed in the next section,
we should similarly ignore such contribution to $\overline{D3}$ for consistency.}.

The canonical normalization for a residual isometry direction $\Theta$ on $S^3$ is given
in Ref.~\cite{D3vacua} such that
\begin{equation}\label{DefCantheta}
\vartheta = \sqrt{T_3g_sM\alpha'} e^{-2u} a_0 \Gamma_3 \Delta\Theta\, .
\end{equation}
To precisely extend the analysis in the near tip region, where the metric is given
by (\ref{eq:TipMetric}), one also should consider the contribution from $S^2$ as the
isometry direction $\Theta$ can generally fiber over both $S^2$ and $S^3$. Assuming
only $\tau$ and $\Theta$ directions are dynamical, the metric takes the generic form
$d\tau^2 + \left( \Gamma_3^2 + \tau^2\Gamma_2^2 \right) d\Theta^2$. 
Finding the canonically normalized residual isometry field would require
diagonalization of such metric. Nevertheless, since $\Gamma_2$ and $\Gamma_3$ can at
most be $\cO(1)$, we expect the canonical normalization (\ref{DefCantheta}) remains
valid at the leading order of a $\tau^2$ expansion. Therefore in such an
approximation, the allowed value for $\vartheta$ at the end of inflation when it
reaches the tachyon surface is bounded by
\begin{equation}\label{Canthetamax}
\vartheta_\fin \leq \vartheta_{\rm c} = \sqrt{T_3g_sM\alpha'} e^{-2u}a_0\Gamma_3
\Delta\Theta_{\rm c} = \frac{a_0 e^{-2u}}{\sqrt{(2\pi)^3 g_s\alpha'}} \, .
\end{equation}
From (\ref{Inflatontau}) and (\ref{tachyonsurface}), we can obtain the relation
between the value of canonical inflaton at the end of inflation and the residual
isometry as
\begin{equation}\label{phiend}
\phit^\fin (\vartheta_\fin) = \sqrt{\frac{T_3}{2} e^{-4u} \left[ a_0^2 \alpha' -
a_0^2 g_sM\alpha'\Gamma_3^2 (\Delta\Theta_\fin)^2 \right]} =
\sqrt{\frac{{\vartheta_{\rm{c}}^2} - {\vartheta^2_\fin}}{2}} \, .
\end{equation}
Until now the analysis has been classical, however additional curvature perturbation
can be generated at the end of inflation due to the quantum fluctuations of
$\vartheta$.

In our case where there are two fields 
{associated with the radial direction which is identified as the inflaton and
the residual isometry direction,}
 following
the $\delta{N}$ formalism~\cite{deltaN} the power spectrum can {\it in principle}
be separated into two parts as
\begin{equation}\label{TotalPowerSpectrum}
\spec_\zeta = \left( \frac{H_k}{2\pi} \right)^2 N_{,\phi}^2 + \left(
\frac{H_k}{2\pi} \right)^2 N_{,\vartheta}^2 \equiv  \spec_{\zeta_\phi} +
\spec_{\zeta_\fin} \, ,
\end{equation}
where the subscript $k$ denotes the quantity evaluated at the moment where the
perturbation associated with the wave number $k$ crosses the horizon during
inflation, and
\begin{equation}
N = \int^{t_{k}}_{t_{\rm e}} H dt
\end{equation}
is the number of $e$-folds. Notice that we have {\em not} used the subscript
{$\phit$ but $\phi$ for the canonical inflaton}, as its definition in terms of the
usual radial coordinate requires precise identification of the horizon exit scale in
the full deformed conifold, and we hope this does not confuse with the angular
coordinate. Here $\spec_{\zeta_\phi}$ is the power spectrum generated by the
canonical inflaton field at the moment of horizon crossing and is given by the
standard formula
\begin{equation}\label{inflatonpowerspectrum}
\spec_{\zeta_\phi} = \frac{1}{2\mpl^2\varepsilon_k} \left( \frac{H_k}{2\pi}
\right)^2 \, ,
\end{equation}
where
\begin{equation}\label{SRepsilon}
\varepsilon \equiv \frac{\mpl^2}{2} \left(
\frac{\partial\mathbb{V}/\partial\phi}{\mathbb{V}} \right)^2
\end{equation}
with $\mathbb{V}$ being the inflaton potential is the slow-roll parameter. Whereas
$\spec_{\zeta_\fin}$ is the additional contribution due to the quantum fluctuations
of $\vartheta$ at the end of inflation, whose explicit form we will write out
shortly. Note that the common prefactor $H_k/(2\pi)$ in (\ref{TotalPowerSpectrum})
comes from the fact that both $\phi$ and $\vartheta$ are relatively light compared
with $H_k$ during inflation.

In general, $\vartheta$ is lighter than the canonical inflaton field (see
(\ref{bulkmass}) and the discussion below) and does not contribute significantly to
the field trajectory. But towards the end of inflation, the isometry direction
$\vartheta$ comes into the play since $\phit^\fin$ does depend on the light field
$\vartheta_\fin$ via (\ref{phiend}) and in turn its quantum fluctuation
$\delta\vartheta({\bf x})$ can give spatial dependence to $\phit^\fin$. In other
words, $\phit^\fin({\bf x})$ takes slightly different values at different parts of
the universe, and such spatial variations can be quantified using the perturbation
in the number of $e$-folds at the end of inflation, $\delta{N}|_\fin =
\zeta_\fin(\mathbf{x})$. This {\em extra} $\zeta_\fin$ at the end of inflation is a
new contribution to the total curvature perturbation other than $\zeta_{{\phi}}$ due
to the canonical inflaton.

Let us now derive the explicit form of $\spec_{\zeta_\fin}$. Since the inflationary
epoch is completely dominated by the canonical inflaton $\phi$, we have the single
field result
\begin{equation}\label{Nprime}
N_{,{\phi}} \equiv \frac{\partial{N}}{\partial{\phi}} = \frac{H}{\dot{{\phi}}}
\end{equation}
so that
\begin{equation}\label{Nprime2}
N_{,{\phi}}^2 = \left( \frac{H}{\dot{{\phi}}} \right)^2 =
\frac{1}{2\mpl^2\varepsilon} \, .
\end{equation}
The derivative of the extra $e$-folds at the final
moment is given by
\begin{equation}\label{endchain}
\left. \frac{\partial{N}}{\partial\vartheta} \right|_\fin = \left.
\frac{\partial{N}}{\partial{\phi}}\frac{\partial{\phi}}{\partial\vartheta}
\right|_\fin = \frac{1}{\sqrt{2\mpl^2\varepsilon_\fin}}
\frac{\partial{\phit}^\fin}{\partial\vartheta_\fin} \, ,
\end{equation}
where the subscript $\phit$ indicates the canonical inflaton near the tip given by
(\ref{Inflatontau}) and the derivative
${\partial{\phit}^\fin}/{\partial\vartheta_\fin}$ can be derived from (\ref{phiend})
as
\begin{equation}\label{dphidtheta}
\frac{\partial\phit^\fin}{\partial\vartheta_\fin} = -
\frac{\vartheta_\fin}{\sqrt{2}\sqrt{\vartheta^2_{\rm c}-\vartheta^2_\fin}} \, .
\end{equation}
Therefore the additional power spectrum generated at the end of inflation is given by
\begin{align}\label{Pend}
\mathcal{P}_{\zeta_\fin} = & \frac{1}{2\mpl^2\varepsilon_\fin} \left[
\frac{\partial{\phit}^{(\fin)}}{\partial\vartheta_\fin} \right]^2 \left(
\frac{H_k}{2\pi} \right)^2
\nonumber\\
= & \frac{1}{4 \varepsilon_\fin}
\frac{\vartheta_\fin^2}{\vartheta^2_\mathrm{c}-\vartheta_\fin^2} \left(
\frac{H_k}{2\pi\mpl} \right)^2 \, .
\end{align}
Here, $\varepsilon_\fin$ is the slow-roll parameter evaluated at $\phit=\phit^\fin$,
and one can substitute away the $\vartheta_\fin$ dependence above using
(\ref{phiend}). For this contribution to dominate, by comparing
(\ref{inflatonpowerspectrum}) with (\ref{Pend}), we require
\begin{equation}
\left| \frac{\partial{\phit}^\fin}{\partial\vartheta_\fin} \right| \gtrsim
\sqrt{\frac{\varepsilon_\fin}{\varepsilon_k}} \, .
\end{equation}
Using (\ref{dphidtheta}) this becomes the condition
\begin{equation}\label{Pcondition}
\vartheta_k \gtrsim \sqrt{\varepsilon_\fin \left(\varepsilon_\fin +
\frac{\varepsilon_k}{2} \right)^{-1}} \vartheta_\mathrm{c} \, ,
\end{equation}
where we have used the fact that $\vartheta$ is very flat so that its
amplitude is almost frozen during the whole inflationary phase, i.e. $\vartheta_k
\sim \vartheta_\fin$.
In order for the two contributions to the power spectrum $\spec_\zeta$ in
(\ref{TotalPowerSpectrum})
to be comparable, we need
the slow-roll parameter  $\varepsilon$ to remain small at the end of inflation, so that
(\ref{dphidtheta}) to be of $\mathcal{O}(1)$.
However if such conditions are satisfied,
as can be read from
(\ref{Pend}), the resulting power spectrum is very sensitive to the angular motion
towards the end of inflation and thus the naive prediction of $\spec_\zeta$ based on the
estimate made far from the tip can be completely spoiled.

To estimate the non-linear parameter $\fnl$~\cite{Komatsu:2001rj}, we need to go
beyond the leading expansion of $\zeta_\fin$: using (\ref{endchain}), we can easily
find that
\begin{equation}
\frac{3}{5}\fnl \approx \frac{1}{2} \left\{ \left.
\frac{\partial^2N/\partial\phi^2}{(\partial{N}/\partial\phi)^2} \right|_\fin +
\frac{\partial^2\phit^\fin/\partial\vartheta^2_\fin}{(\partial{N}/\partial\phi|_\fin)
[\partial\phit^\fin/\partial\vartheta_\fin]^2} \right\} \, .
\end{equation}
From (\ref{Nprime}), we can see that the first term in the curly brackets becomes
\begin{equation}
\left. \frac{\partial^2N/\partial\phi^2}{(\partial{N}/\partial\phi)^2} \right|_\fin
= \eta_\fin - 2\varepsilon_\fin \, ,
\end{equation}
where
\begin{equation}\label{SReta}
\eta \equiv \mpl^2 \frac{\partial^2\mathbb{V}/\partial\phi^2}{\mathbb{V}}
\end{equation}
is another slow-roll parameter, and the second term
\begin{equation}
\frac{\partial^2\phit^\fin/\partial\vartheta^2_\fin}{(\partial{N}/\partial\phi|_\fin)
[\partial\phit^\fin/\partial\vartheta_\fin]^2} = 2\sqrt{\varepsilon_\fin} \left(
\frac{\vartheta_\mathrm{c}}{\vartheta_\fin} \right)^2
\frac{\mpl}{\sqrt{\vartheta_\mathrm{c}^2 - \vartheta^2_\fin}} \, .
\end{equation}

So far we have only considered the simplified situation where only the tachyon
potential (\ref{Tachyonpotential}) is present and have hence ignored other potential
terms which can also become dominant near the end of inflation. One candidate is the
$D3$-$\overline{D3}$ Coulombic interaction in (\ref{DyD3barD3}) which can be ignored at
 large radius where the singular conifold
approximation is sufficient, but can dominate near the tip of the deformed conifold. In
fact as both the Coulombic and the tachyon potential depend on the
$D3$-$\overline{D3}$ separation $\sepy$, {if they dominate towards the end of inflation},
 the inflationary trajectory would be driven
to incident on the tachyon surface at a right angle. By an appropriate rotation in
the $\phit$-$\vartheta$ plane, the effect described earlier can then be shown to
vanish, as on the tachyon surface there is no orthogonal component for the field
trajectory. To have a significant Lyth effect as we described above, it is {\em
necessary} in our case to ensure that the Coulombic potential is insignificant,
hence the end-of-inflation surface differs from the equi-energy
surface\footnote{This is however not necessarily true in general, since the number
of $e$-folds depends {\em non-locally} on the dynamics during inflation: see e.g.
Ref.~\cite{endtraj}. We thank Misao Sasaki for related communications.}.

Another necessary criterion for the Lyth effect to give a significant contribution
is that the slow-roll parameter $\varepsilon$ remains small at the onset of
tachyon condensation: in other words, inflation should persist into the deformed
conifold region. As we will demonstrate explicitly in the later sections and
appendices, the Coulombic interaction which tend to give large $\varepsilon$ near
the tip can be naturally made insignificant (depending on the uplifting mechanisms), and so
inflation ends only when the $D3$-$\overline{D3}$ annihilates.

\section{An alternative scheme for uplifting}
\label{sec_scenario}
\setcounter{equation}{0}
\paragraph{}

Having reviewed the Lyth effect and the necessary conditions for it to take place in
brane inflation, in this section, we will begin with elucidating different possible
uplifting mechanisms for generating de Sitter vacua necessary for a realistic vacua
at the end of inflation. By considering the relative strengths between these
potentials in a warped throat, we then propose an alternative scenario where the
distant sources or $D$-terms on $D7$ branes dominate over the $\overline{D3}$ at the
tip of deformed conifold in contributing to the vacuum energy, and are responsible
for the majority of uplifting. This allows us to decouple the $D3$-$\overline{D3}$
Coulombic potential {towards the end of inflation.}

\subsection{Uplifting potentials}
\paragraph{}

In general, the $F$-term scalar potential $V_F$ generated by flux and
non-perturbative correction gives rise to an anti de Sitter minimum after all the
moduli are stabilized~\cite{KKLT}. To obtain a de Sitter vacuum at the end of
inflation, it is therefore necessary to include extra uplifting term(s) to raise the
cosmological constant to a positive value. In the setup described earlier, the
leading term $D_0/[U(r,\sigma)]^2$ in the $D3$-$\overline{D3}$ potential given by
(\ref{DefVD3barD3}) essentially plays that role. To obtain a small positive
cosmological constant, one can estimate that at the tip of the deformed
conifold~\cite{Baumann1}
\begin{equation}
1< \frac{D_0/[U(\epsilon^{2/3},\sigma_F)]^2}{|V_F(\epsilon^{2/3},\sigma_F)|}
\lesssim {\mathcal{O}}(3) \, .
\label{coupledVFVD}
\end{equation}
Here $\sigma_F$ is the stabilized volume before the uplifting and $\epsilon^{2/3}$
indicates that the potentials are evaluated at the bottom of the throat.
One should also note that the upper bound is required for the stability of the
$\sigma_F$.
The requirement (\ref{coupledVFVD}) couples the scale of the Coulombic potential $D_0$
to the scale of $V_F$. Away from the tip, the adiabatic approximation can be taken
such that
 $\sigma$ remains at its
instantaneous minimum at each radial location, and $\sigma_\star(r)$ can be shown
to be a monotonously
increasing function of $r$. Since $|V_F|\sim \exp(-a\sigma)/\sigma$ which reaches
its maximum at the tip, one can then ensure a positive cosmological constant
provided that the lower bound of (\ref{coupledVFVD}) is satisfied.

In addition, if there are also distant $\overline{D3}$ branes present outside the
throat, for example in other throat(s), they can also contribute to the vacuum
energy and their contribution can be given by
\begin{equation}\label{DefVother}
V_{\rm other} = \frac{D_{\rm other}}{[U(r,\sigma)]^2} \, .
\end{equation}
In general, we do not know the explicit value of $D_{\rm other}$. However it is
important to know that as these extra $\overline{D3}$ branes are outside the warped
deformed conifold, $D_{\rm other}$ is independent of the warp factor $a_0$, and such
a contributions\ can outweigh $V_{D3\overline{D3}}$ whose magnitude is controlled by
$a_0$. Inclusion of such a contribution will also modify (\ref{coupledVFVD}) from
$D_0\to D_0+D_{\rm other}$.

An alternative approach was suggested in Ref.~\cite{BKQ} by localizing supersymmetry
breaking flux on the $D7$ brane worldvolume\footnote{The four cycle where $D7$ wraps
on does not necessarily have to be the ones where gaugino condensation takes place.},
which induces a $D$-term potential in the low energy effective four dimensional
supergravity. The advantage of this approach is that the uplifting effect can be
studied within a field theoretical framework. For our purpose, in the most
simplistic setup\footnote{By simplistic we means that we have ignore the
contribution coming from the additional matter field charged under the $U(1)$ gauge
field associated with ${\mathcal{F}}_{D7}$, and furthermore this configuration can
be generalized to non-Abelian gauge group $U(N)$.}, such $D$-term potential is given
schematically by
\begin{equation}\label{VDterm1}
V_{D\mathrm{-term}}(r,\sigma) = \frac{v_D}{[U(r,\sigma)]^3} \, .
\end{equation}
The precise value of the constant $v_D$ depends on the explicit world
volume flux ${\mathcal{F}}_{D7}$ and is proportional to the integral
$\int_{\Sigma_4} \hat{J}\wedge {\mathcal{F}}_{D7}$~\cite{D7Dterm}, where $\hat{J}$
is the pull-back of the K\"ahler form of the ambient Calabi-Yau onto the four cycle the
$D7$ wraps on. The four cycles which $D7$ branes wrap on can be outside the warped
throat or if they are inside the warp throat, explicit power counting can then show
that $v_D$ should contain additional extra warped factor $a_0^4$~\cite{Dterm3}.

Notice that while (\ref{VDterm1}) is proportional to $[U(r,\sigma)]^{-3}$, unlike
$V_{D3\overline{D3}}$, it does not depends on the $D3$-$\overline{D3}$ separation
$|\mathbf{y}-\mathbf{\bar{y}}|$. Furthermore, as noticed in
Refs.~\cite{Dterm2,Mirage} and explicitly demonstrated in Ref.~\cite{Dterm3} (using
the results of Ref.~\cite{Dterm1}) $D$-term uplifting is subjected to an extra
constraint. In $\mathcal{N}=1$ supergravity, the magnitude of the $D$-term potential
is in fact proportional to that of $F$-term $D_{\Sigma} W$, therefore the $D$-term
potential (\ref{VDterm1}) cannot uplift a supersymmetric anti de Sitter minimum
satisfying $D_{\Sigma}V_F=0$\footnote{This is a generic feature of $V_F$ of KKLT
type, however such a $D$-term has been shown to uplift non-supersymmetric anti de
Sitter minimum~\cite{Dterm3,Mirage}.}. However by explicitly introducing
$\overline{D3}$ hence breaking supersymmetry, we can in principle circumvent such
constraint\footnote{We are grateful to Fernando Quevedo for a discussion on this point.}.

We can write these uplifting potentials in an universal fashion as
\begin{equation}
V_{D}(r,\sigma) = \frac{D(|\mathbf{y}-\mathbf{\bar{y}}|)}{[U(r,\sigma)]^b} \, ,
\end{equation}
where
\begin{equation}\label{DefVDuplifting}
D(|\mathbf{y}-\mathbf{\bar{y}}|) = \left\{
\begin{split}
 & D_0 \left( 1 - \frac{3D_0}{16\pi^2T_3^2|\mathbf{y}-\mathbf{\bar{y}}|^4} \right) +
 D_\mathrm{other} & \mbox{for } D3\mbox{-}\overline{D3} & ~~~(b=2) \, ,
 \\
 & v_D & \mbox{for } D\mbox{-term} & ~~~(b=3) \, .
\end{split}
\right.
\end{equation}
Here we have included the term
\begin{equation}\label{DefVcoul}
V_{\rm Coulomb} = -\frac{D_0}{[U(r,\sigma)]^2}
\frac{3D_0}{16\pi^2T_3^2|\mathbf{y}-\mathbf{\bar{y}}|^4}
\end{equation}
in the $D3$-$\overline{D3}$ interaction to highlight the fact that its scale is also
set by $D_0$, even though it gives a negative contribution to the total energy. The
interplay between the $D$-term potential $V_D$ and the $F$-term scalar potential
$V_F$ will become crucial when we later consider the possibility of generating
significant contribution to the curvature perturbation at the end of inflation.

\subsection{Proposed scenario}
\paragraph{}

In contrast to Ref.~\cite{Baumann1} which we briefly review in Appendix
\ref{app_Baumann}, in the scenario we will consider, while a $\overline{D3}$ brane
can still be present at the tip of the deformed conifold for tachyon condensation to
take place at the end of the inflation, the additional distant $\overline{D3}$
branes or supersymmetry breaking $D7$ branes will be responsible for uplifting. That
is, in terms of their magnitude,
\begin{equation}\label{Decoupling1}
D_{\rm other}\gg D_0 \, ,
\end{equation}
or equivalently
\begin{equation}\label{Decoupling2}
v_D\gg D_0 U(r,\sigma) \, .
\end{equation}
In other words, we would like to decouple the $D_0$ dependent terms in
(\ref{DefVDuplifting}): in particular the Coulombic term  $V_{\rm Coulomb}\sim
-D_0^2|\mathbf{y}-\mathbf{\bar{y}}|^{-4} \propto a_0^8
|\mathbf{y}-\mathbf{\bar{y}}|^{-4}$ is decoupled in the entire throat.

Such decoupling of Coulombic interaction is very natural. The maximal
magnitude $V_{\rm Coulomb}$ can take is given by $3 a_0^4/(4\pi^2{\alpha'}^2U^2)$,
which corresponds to the $D3$-$\overline{D3}$ separation
$|\mathbf{y}-\mathbf{\bar{y}}|^2$ on the tachyon condensation surface. Without
additional dominating uplifting sources, such potential dominates near the tip region,
%
and the scale of $|V_F(\epsilon^{2/3},\sigma_F)|$ is therefore coupled to that of the
Coulombic term $D_0$. This also implies  $|V_F(\epsilon^{2/3},\sigma)|\propto
a_0^4$, although $V_F$ does not contain $a_0$ in its expression {\it a priori}. However in
the presence of additional dominating uplifting sources, the scale of $V_F$ does not
have to couple to $D_0$, but rather should couple to these additional uplifting
terms whose magnitudes are independent of the warp factor $a_0$. This can allow
$V_F(r,\sigma)-V_F(\epsilon^{2/3},\sigma_F)$ to dominate over the
$D3$-$\overline{D3}$ Coulombic interaction $V_{\rm Coulomb}$ not only at large
radius but also in the near tip region.

Here we introduce a parameter $s$ given by the ratio
\begin{equation}\label{Defs}
s = \frac{V^{(+)}_D(\epsilon^{2/3},\sigma_F)}{|V_F(\epsilon^{2/3},\sigma_F)|} \, ,
\end{equation}
where $V^{(+)}_D$ denotes that we are only keeping the positive definite term in
both expressions in (\ref{DefVDuplifting}).
This allows us to write the overall potential we are considering schematically as
\begin{align}\label{Schempotential}
\mathbb{V} = & V_F(r,\sigma) + V_D(r,\sigma) + V_{\rm end}
\nonumber\\
= & \delta{V_F} + \delta{V_D^{(+)}} + V_{\rm Coulomb}+V_{\rm end} + (s-1) \left|
V_F(\epsilon^{2/3},\sigma_F) \right| \, ,
\end{align}
where
\begin{align}
\delta{V_F} = & V_F(r,\sigma) - V_F(\epsilon^{2/3},\sigma_F) \, ,
\\
\delta{V_D^{(+)}} = & V_D^{(+)}(r,\sigma) - V_D^{(+)}(\epsilon^{2/3},\sigma_F) \, ,
\end{align}
and $V_{\rm end}$ consists of the potentials that only becomes significant near the
end of inflation, e.g. the tachyon potential (\ref{Tachyonpotential}) and the
possible bulk mass term for the residual isometry direction. Notice that as
$U(r,\sigma)$ can be shown to be a monotonously increasing function of $r$,
$\delta{V_F}$ is positive definite while $\delta{V^{(+)}_D}$ is negative definite.
With a slight abuse of notation, here we have not specified $V_D$: it can in
principle consist of contributions from the distant $\overline{D3}$, or
supersymmetry breaking $D7$, or both along with $\overline{D3}$ at the tip. The
condition (\ref{Decoupling1}), or equivalently (\ref{Decoupling2}), then translates
into the requirement
\begin{equation}\label{Decoupling3}
\delta{V_F} \gg V_{\rm Coulomb} \, ,
\end{equation}
and we consider the situation where this condition holds for all values of the
mobile $D3$ brane coordinates. In terms of the available parameters which we can
tune, (\ref{Decoupling3}) translates into the condition $|A_0|\mpl^{-3} \gg
a_0^4/(\mpl^4{\alpha'}^2)$, where $|A_0|$ appears in (\ref{eq:SuperW}). As
$\mpl^2\alpha'\ge 1$ and $a_0\ll 1$, the condition (\ref{Decoupling3}) can be easily
met with suitable choice of $A_0$.

In the absence of $D$-term uplifting potential, such decoupling of $V_{\rm
Coulomb}\sim a_0^8\sepy^{-4}$ does not yield significant qualitative differences to
the overall inflaton potential at large radius $r\gg \epsilon^{2/3}$. The canonical
inflation potential should behave qualitatively similar to the one in
Ref.~\cite{Baumann1} (see (\ref{delicatepotential})). In fact, one can show that the
potential (\ref{delicatepotential}) can yield small slow-roll parameter
$\varepsilon$ until very small radius (see Appendix \ref{app_Baumann}), that is,
inflation can persist well into the deformed conifold. Moreover, at small radius
$r\approx \epsilon^{2/3}$ where inflation ends, the condition (\ref{Decoupling3})
can in principle allow for significant contributions to the curvature perturbation
via the Lyth effect discussed earlier, leading to noticeable changes in the power
spectrum $\spec_\zeta$ and the non-linear parameter $\fnl$ due to the residual
isometry direction.

\section{An explicit case study of the Lyth effect in brane inflation}
\setcounter{equation}{0}
\label{sec_endofinf}
\paragraph{}

In this section, we will first calculate the canonical inflaton potential near the
tip of the deformed conifold with non-perturbative superpotential generated by the
Kuperstein embedding~\cite{Kuper}, and demonstrate that the slow-roll parameter
$\varepsilon$ can remain small near the tip region for the uplifting scenario
described in the previous section. We will then discuss the possibility of the Lyth
effect in this setup. We will demonstrate that, for the specific angular stable
trajectory of this embedding, the residual isometry direction becomes degenerate for
the entire deformed conifold, hence the accidental disappearance of the Lyth effect.
We therefore conclude that while the general setup described earlier constitutes the
necessary criteria for the residual isometries to significantly affect observations,
the angular stable inflationary trajectory, governed by the geometry of the specific
embedding, will determine whether it {\it actually} takes place or not.

\subsection{Potential near the tip of deformed conifold}
\paragraph{}

Near the tip of the deformed conifold, the complicated K\"ahler potential
(\ref{DefKahdefcon}) simplifies to (\ref{asymptotic_k}) after using the constraint
(\ref{Defepsilon}) to rewrite
\begin{equation}
z^4 = \pm\left[ \epsilon^2 - \sum_{i=1}^3(z^i)^2 \right]^{1/2} \, .
\end{equation}
Using the formula given in Ref.~\cite{McGill1}, the metric and its inverse derived
from the simplified K\"ahler potential (\ref{asymptotic_k}) are given by
\begin{align}
\label{NTmetric}
k_{i\bar{j}} = & \frac{c}{\epsilon^{2/3}} \left( \delta_{i\bar{j}} +
\frac{z_i\bar{z}_j}{|z^4|^2} \right) \, ,
\\
\label{NTImetric}
k^{\bar{i}{j}} = & \frac{\epsilon^{2/3}}{c} \left( \delta^{\bar{i}{j}} -
\frac{z^i\bar{z}^j}{r^3}\right) \, .
\end{align}
Here the indices $\bar{i}$, ${j} = 1, 2, 3$, where we have also used 
(\ref{Defr}). Raising and lowering of the indices is done by $\delta^i{}_j$. Using
(\ref{NTmetric}) and (\ref{NTImetric}), we can find the $F$-term scalar potential
valid near the tip of deformed conifold as
\begin{equation}\label{VFTip}
V_F = V_\mathrm{KKLT} + \Delta{V}_F \, ,
\end{equation}
where
\begin{align}
\label{DefVKKLT}
V_\mathrm{KKLT} = & \frac{2\kappa^2 a e^{-2a\sigma}|A(z^i)|^2}{[U(r,\sigma)]^2}
\left\{ 1 + W_0 e^{a\sigma} {\rm Re} \left[ \frac{e^{ia\varsigma}}{A(z^i)} \right] +
\frac{a}{6} \left[ 2\sigma - \gamma k_0 + c\gamma\epsilon^{4/3} \left( 1 -
\frac{\epsilon^2}{r^3} \right) \right] \right\} \, ,
\\
\label{DefDeltaVF}
\Delta{V}_F = & \frac{\kappa^2 e^{-2a\sigma}}{3[U(r,\sigma)]^2} \left\{
\frac{\epsilon^{2/3}}{c\gamma} \left[ |A_i|^2 - \frac{\left(
\sum_{i=1}^3\bar{z}^iA_i \right) \left( \sum_{j=1}^{3}z^j\overline{A}_{\bar{j}}
\right)}{r^3} \right] - {\rm Re} \left[ \overline{A}A_i \left( z^i - \bar{z}^i
\frac{\epsilon^2}{r^3} \right) \right] \right\} \, ,
\end{align}
where $A_i = \partial A(z)/\partial z^i$ and $\overline{A}_{\bar{j}} =
\partial\overline{A(z)}/\partial\bar{z}^j$ so that if $A(z^i)$ becomes constant
$\Delta V_F$ vanishes and $V_{\rm KKLT}$ reduces to the $F$-term scalar potential
considered in Refs.~\cite{KKLMMT,KKLT}. Here we have separated the
contribution from the dependence of non-perturbative superpotential on the mobile
$D3$ brane,  $\Delta V_F$.

We can now again consider specifically the Kuperstein embedding~\cite{Kuper}
\begin{equation}\label{Kuperemb1}
f(z^i)=z^1-\mu \, ,
\end{equation}
and without loss of generality we will take real deformation parameter $\mu\in
{\mathbb{R}}$ and $\epsilon\in{\mathbb{R}}$ as noted earlier. The function $A(z^j)$
and $A_i(z^j)$ in (\ref{DefVKKLT}) and (\ref{DefDeltaVF}) become
\begin{align}
\label{Az}
A(z^j)= & A_0\left(1-\frac{z^1}{\mu}\right)^{1/n} \, ,
\\
\label{Ai}
A_i(z^j) = & -\frac{A_0}{n\mu} \left( 1 - \frac{z^1}{\mu} \right)^{1/n-1}
\delta_{i1} \, .
\end{align}
Evidently $A(z^1)$ and $A_i(z^1)$ should preserve an $SO(3)$ residual symmetry group
of rotation among $\{ z^2,z^3,z^4 \}$. Importantly, with such a choice of $D7$ embedding, the
$F$-term scalar potential again reduces to a function of $\left\{\sigma, \varsigma,
r, |z^1|^2, z^1+\bar{z}^1 \right\}$ instead of all deformed conifold coordinates,
i.e.
\begin{equation}\label{SimpleVF}
V_F \equiv V_F(\sigma, \varsigma, r, |z^1|^2, z^1+\bar{z}^1) \, .
\end{equation}
The angular coordinates that appear explicitly in $V_F(\sigma, \varsigma, r,
|z^1|^2, z^1+\bar{z}^1)$ correspond to the broken isometry directions and they are
exclusively encoded in the combinations $|z^1|^2$ and $z^1+\bar{z}^1$, therefore to
obtain the angular extremum trajectory amounts to finding the trajectory where
\begin{equation}\label{AngStabCond}
\frac{\partial |z^1|^2}{\partial \Psi_i}=\frac{\partial (z^1+\bar{z}^1)}{\partial
\Psi_i}=0 \, ,
\end{equation}
where $\{ \Psi_i \}$ include all the broken angular isometry directions of the
deformed conifold. In Appendix \ref{app_stability}, we explicit obtain the angular
stable trajectory given by
\begin{align}\label{AngStabTraj2_z1}
z^1 = & -\sqrt{\frac{r^3+\epsilon^2}{2}} \, ,
\\\label{AngStabTraj2_z2}
z^2 = & \pm i\sqrt{\frac{r^3-\epsilon^2}{2}} \, ,
\\\label{AngStabTraj2_rest}
z^3 = & z^4 =  0 \, .
\end{align}
Along such trajectory, the $SO(3)$ residual isometry preserved by the $D7$ embedding
(\ref{Kuperemb1}) is further broken down to $SO(2)$ rotating $z^3$ and $z^4$. We can
also stabilize the axion field $\varsigma$ as in Ref.~\cite{Baumann1}, by arranging
$W_0$ to be a small negative constant. The resultant two-field potential is then
given by $V_F(r,\sigma) = V_\mathrm{KKLT}(r,\sigma) + \Delta{V}_F(r,\sigma)$, where
\begin{align}
\label{VKKLT2}
V_\mathrm{KKLT}(r,\sigma) = & \frac{2\kappa^2 a e^{-2a\sigma}
|A_0|^2}{[U(r,\sigma)]^2} \left( 1 + \frac{\sqrt{r^3+\epsilon^2}}{\sqrt{2}\mu}
\right)^{2/n}
\nonumber\\
& \times \left\{ 1 - e^{a\sigma} \frac{|W_0|}{|A_0|} \left( 1 +
\frac{\sqrt{r^3+\epsilon^2}}{\sqrt{2}\mu} \right)^{-1/n} + \frac{a}{6} \left[
2\sigma-\gamma k_0 + \gamma c \epsilon^{4/3} \left( 1 - \frac{\epsilon^2}{r^3} \right)
\right] \right\} \, ,
\\
\label{DeltaVF2}
\Delta V_{F}(r,\sigma) = &
\frac{\kappa^2e^{-2a\sigma}|A_0|^2}{3n^2[U(r,\sigma)]^2} \left( 1 +
\frac{\sqrt{r^3+\epsilon^2}}{\sqrt{2}\mu} \right)^{2(1/n-1)} \left( 1 -
\frac{\epsilon^2}{r^3} \right)
\nonumber\\& \times
\left[ \frac{\epsilon^{2/3}}{2\mu^2c\gamma} - 2an
\frac{\sqrt{r^3+\epsilon^2}}{\sqrt{2}\mu} \left( 1 +
\frac{\sqrt{r^3+\epsilon^2}}{\sqrt{2}\mu} \right) \right] \, .
\end{align}
We will now include the effect of the uplifting potential as given in
(\ref{DefVDuplifting}). In our scenario, we have decoupled $V_{\rm Coulomb}$, so we
should strictly include the positive definite term, i.e. $V_D^{(+)}$. We can
also further integrate out $\sigma$ by assuming that $\sigma$ evolves adiabatically
and remains at its instantaneous minimum, which is given by
\begin{equation}\label{Defsigmastar}
\left. \frac{\partial \left[ V_F+V_D^{(+)}
\right](r,\sigma)}{\partial\sigma}\right|_{\sigma = \sigma_\star(r)} = 0 \, .
\end{equation}
This eventually leads to a single field potential
\begin{equation}\label{BFVr}
\mathbb{V}(r) = V_{F}[r,\sigma_\star(r)] + V_D^{(+)}[r,\sigma_\star(r)] + V_{\rm
end} \, .
\end{equation}
Notice that we have not specified whether $V^{(+)}_D$ is attributed to distant
$\overline{D3}$ or $D7$, as in the absence of $V_{\rm Coulomb}$, these two cases can
be treated on equal footing calculationally. (\ref{Defsigmastar}) is in fact a
transcendental equation, which is solved numerically in general. But in Appendix
\ref{app_vol} we derive the lowest order approximated expression given by
\begin{equation}\label{Mainapproxvolume2}
\sigma_\star(r) \approx \sigma_0 \left[ 1 + \frac{c_1}{a\sigma_0}(r-\epsilon^{2/3})
\right] \, ,
\end{equation}
where the coefficient $c_1$ is given by
\begin{equation}\label{C1}
c_1 = \frac{3\varepsilon^{1/3}}{4n\mu} \left( 1 + \frac{\varepsilon}{\mu}
\right)^{-1} + \mathcal{O} \left( \frac{1}{\sigma_0} \right) \, .
\end{equation}
Finally, we note that the function $r(\phit)$ can be derived from (\ref{Defr}) and
(\ref{Inflatontau}), and a good working expression relating canonical inflaton to
the radial coordinate $r$ in this region can be given by
\begin{equation}\label{Defrphit}
r(\phit) = \sqrt{\frac{2}{3T_3}\phit^2 + \epsilon^{4/3}} \, .
\end{equation}

Now we have all the information to write the single field inflaton potential near
the tip of the deformed conifold. Putting (\ref{VKKLT2}), (\ref{DeltaVF2}),
(\ref{Mainapproxvolume2}) and (\ref{Defrphit}) together, the single field potential
for the canonical inflaton $\phit$ along the angular stable trajectory
$z^1=-\sqrt{(r^3+\epsilon^2)/2}$ is given by
\begin{equation}\label{MathbbVphi}
\mathbb{V}(\phit) = V_\mathrm{KKLT} \left[ r(\phit),\sigma_\star(\phit) \right] +
\Delta{V}_{F} \left[ r(\phit),\sigma_\star(\phit) \right] + V_D^{(+)} \left[
r(\phit),\sigma_\star(\phit) \right] + V_{\rm end} \, ,
\end{equation}
where
\begin{align}
V_\mathrm{KKLT}[r,\sigma_\star(r)] = &
\frac{2\kappa^2a|A_0|^2e^{-2a\sigma_\star(r)}}{\{U[r,\sigma_\star(r)]\}^2} \left( 1
+ \frac{\sqrt{r^3+\epsilon^2}}{\sqrt{2}\mu} \right)^{2/n}
\nonumber\\
& \times \left\{ 1 - \frac{|W_0|}{|A_0|}e^{a\sigma_\star(r)} \left( 1 +
\frac{\sqrt{r^3+\epsilon^2}}{\sqrt{2}\mu} \right)^{-1/n} + \frac{a}{6} \left[
2\sigma_\star(r) - \gamma{k_0} + \gamma{c}\epsilon^{4/3} \left( 1 -
\frac{\epsilon^2}{r^3} \right) \right] \right\} \, ,
\label{VKKLTsingle}\\
\Delta{V_F}[r,\sigma_\star(r)] = &
\frac{\kappa^2|A_0|^2e^{-2a\sigma_\star(r)}}{3n^2\{U[r,\sigma_\star(r)]\}^2} \left(
1 + \frac{\sqrt{r^3+\epsilon^2}}{\sqrt{2}\mu} \right)^{2(1/n-1)} \left( 1 -
\frac{\epsilon^2}{r^3} \right)
\nonumber\\
& \times \left[ \frac{\epsilon^{2/3}}{2c\mu^2\gamma} - 2an
\frac{\sqrt{r^3+\epsilon^2}}{\sqrt{2}\mu} \left( 1 +
\frac{\sqrt{r^3+\epsilon^2}}{\sqrt{2}\mu} \right) \right] \, ,
\label{DeltaVFsingle}\\
V_D^{(+)}[r,\sigma_\star(r)] = & \frac{D^{(+)}(\sepy)}{\{U[r,\sigma_\star(r)]\}^b}
\label{VDsingle}\,.
\end{align}
The function $D^{(+)}(\sepy)$ can be read off from (\ref{DefVDuplifting}) by keeping
only positive definite term and $V_{\rm end}$ consists of the potentials that are
significant at the end of inflation, e.g. the tachyon potential.

\subsection{Slow-roll parameter near the tip of the throat}
\paragraph{}

\begin{figure}[h]
\begin{center}
\includegraphics[width = 8cm]{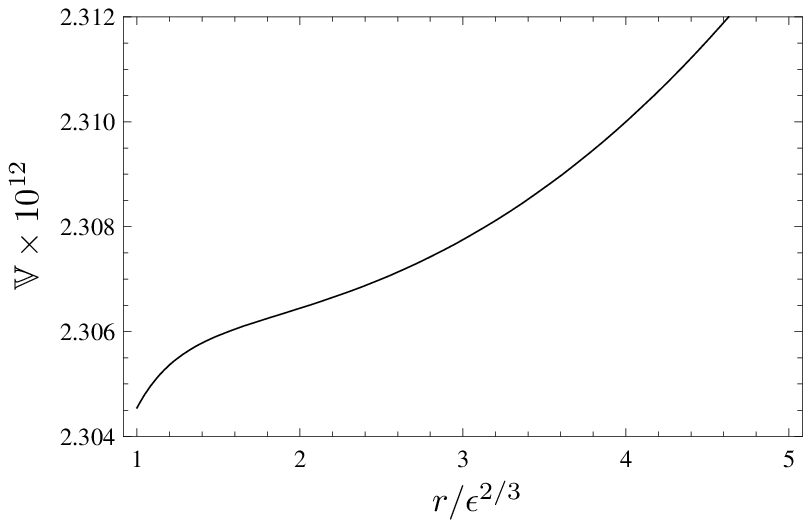}
\includegraphics[width = 8cm]{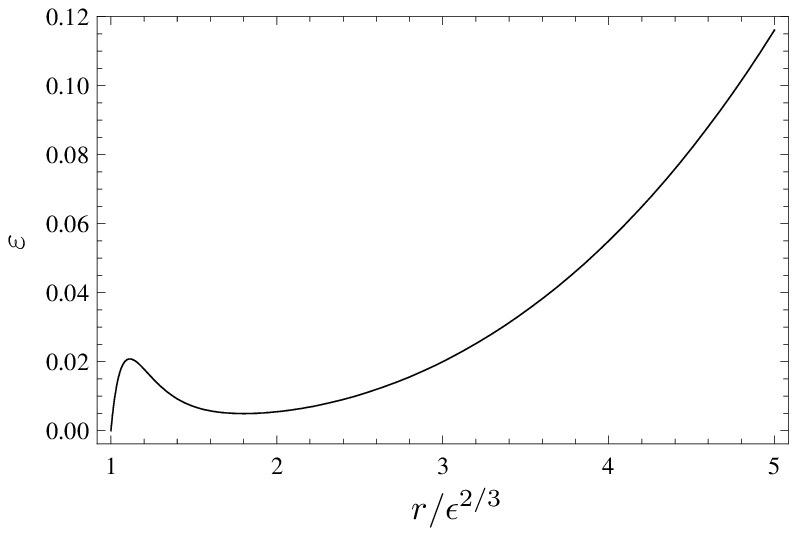}
\end{center}
\caption{(Left) the inflaton potential (\ref{MathbbVphi}) and (right) the slow-roll
parameter $\varepsilon$ (\ref{epsilonSR2}). {We normalize $\mpl = 1$ and for
simplicity we set $A_0 = 1$.} The point $r/\epsilon^{2/3} = 1$ denotes the tip from
which there is no further radial displacement. As shown in the right panel, the
potential is very flat near the tip.}
 \label{tipplot}
\end{figure}

Given the inflaton potential as (\ref{MathbbVphi}), now we can calculate the
slow-roll parameter (\ref{SRepsilon}) which is needed in determining the overall
amplitude of the power spectrum $\spec_\zeta$ as (\ref{Pend}) and we will
demonstrate that it can remain small near the tip of deformed conifold in our
scenario, i.e. the inflaton potential (\ref{MathbbVphi}) is very flat near the tip.
By chain rule, we can write
\begin{align}\label{epsilonSR2}
\varepsilon = & \frac{\mpl^2}{2} \left( \frac{\partial{r}}{\partial\phit} \right)^2
\left( \frac{\partial\mathbb{V}/\partial{r}}{\mathbb{V}} \right)^2 \, ,
\end{align}
where, using (\ref{Defrphit}), the derivative of $r$ with respect to $\phit$ is
given by
\begin{align}\label{dphitdrs}
\frac{\partial{r}}{\partial\phit} =  \sqrt{\frac{2}{3T_3} \left( 1 -
\frac{\epsilon^{4/3}}{r^2} \right)} \, .
\end{align}
As shown in more detail in Appendix \ref{app_slow-roll}, $\varepsilon$ is a
complicated function of $r$. To get a clearer idea, let us evaluate the expression
(\ref{epsilonSR2}) at the tip: it reads
\begin{align}\label{dVdrVtip}
\left. \frac{\partial{\mbV}/\partial{r}}{\mbV} \right|_{r = \epsilon^{2/3}} = &
\frac{1}{s-1} \left\{ \frac{3-sb}{U(\epsilon^{2/3},\sigma_F)} \left[
\frac{3\epsilon^{1/3}}{4\pi\mu} \cG - c\gamma\epsilon^{2/3} \right] +
\frac{2sbc\gamma}{U(\epsilon^{2/3},\sigma_F)}\epsilon^{2/3} + \frac{s}{D} \left.
\frac{\partial{D}}{\partial{r}} \right|_{r=\epsilon^{2/3}} \right.
\nonumber\\
& \hspace{1.2cm} \left. + \frac{3\cG^2}{4\pi^2U(\epsilon^{2/3},\sigma_F)} \left[
\frac{\epsilon^{2/3}}{2c\mu^2\gamma} - 4\pi \frac{\epsilon}{\mu} \cG^{-1}
\right]\epsilon^{2/3} \right\} \, ,
\end{align}
where
\begin{equation}\label{curlyG}
\cG = \left( 1 + \frac{\epsilon}{\mu} \right)^{-1} \, .
\end{equation}
To work out the numerical value for (\ref{dVdrVtip}), it is useful to express in
terms of the geometric parameters describing the bulk and the throat. How to write
(\ref{dVdrVtip}) in terms of which parameters is described in Appendix
\ref{app_slow-roll}, and the result is
\begin{align}
\left. \frac{\partial{\mbV}/\partial{r}}{\mbV} \right|_{r = \epsilon^{2/3}} = &
\frac{\mu^{-2/3}}{s-1} \left\{ \frac{3-sb}{3NB_4\log{Q_\mu}} \left[ \frac{3}{2}
3^{1/12} \left( \frac{a_0Q_\mu}{c} \right)^{1/2} \left(1 + 3^{1/4} \left(
\frac{a_0Q_\mu}{c} \right)^{3/2} \right) \right.\right.
\nonumber\\
& \left. \hspace{4cm} - \frac{B_4}{B_6} \frac{2\cdot2^{1/3}c\log{Q_\mu}}{3Q_\mu^2}
3^{1/12}\left( \frac{a_0Q_\mu}{c} \right) \right]
+ \frac{4\cdot2^{1/3}sbc}{9NB_6Q_\mu^2} 3^{1/6} \left( \frac{a_0Q_\mu}{c} \right)
\nonumber\\
& \hspace{1.2cm} + \frac{2}{NB_4\log{Q_\mu}}\left[ 1 + 3^{1/4}\left(
\frac{a_0Q_\mu}{c} \right)^{3/2} \right]^{-2}
\nonumber\\
& \left. \hspace{1.5cm} \times \left[ \frac{B_6}{B_4}
\frac{3Q_\mu^2}{8\cdot2^{1/3}c\log{Q_\mu}} - 3^{1/12} \left( \frac{a_0Q_\mu}{c}
\right)^{1/2} \left( 1 + 3^{1/4}\left( \frac{a_0Q_\mu}{c} \right)^{3/2} \right)
\right] \right\} \, .
\end{align}
To obtain a definite number, we use the sample set of parameters given in
Ref.~\cite{Baumann1}: $N = 32$, $Q_\mu = 1.2$, $B_4 =9$, and $B_6 = 1.5$. Then we
can see that to the lowest order expansion around the tip
\begin{equation}\label{epsilonSRtip}
\varepsilon(r) \approx \frac{0.00504265}{(s-1)^2} \left( 1 - \frac{\epsilon^{4/3}}{r^2}
\right) \, .
\end{equation}
Two comments are in order: first, it is clear that exactly at the tip, i.e. $r =
\epsilon^{2/3}$ the slow-roll parameter is simply $\varepsilon_\mathrm{tip} = 0$.
 Second, away from the tip, $\varepsilon(r)$ can be
 reasonably small
 by choosing the parameters to allow
for significant curvature perturbation spectrum at the end of inflation through the
Lyth effect described in Section \ref{Lytheffect}. In Fig.~\ref{tipplot}, we show
the inflaton potential (\ref{MathbbVphi}) and the slow-roll parameter $\varepsilon$
(\ref{epsilonSR2}).

\subsection{The angular stable trajectory and the degenerate residual isometry}
\label{DegAngle}
\paragraph{}

In this section, we will argue that the angular stable trajectory near the tip
region (\ref{AngStabTraj2_z1})-(\ref{AngStabTraj2_rest}) for the Kuperstein
embedding (\ref{Kuperemb1}) is in fact valid along the {\em entire} deformed
conifold by showing explicitly that the extremal values of the broken isometry
directions along (\ref{AngStabTraj2_z1})-(\ref{AngStabTraj2_rest}) are identical to
the ones for the stable trajectory in the singular conifold (\ref{AngStaTra1}),
despite very different scalar potential in each region. Along this specific
trajectory, we will then show that the proper distance associated with the residual
isometry direction preserved by the Kuperstein embedding vanishes.

To begin with, we can rewrite the trajectory
(\ref{AngStabTraj2_z1})-(\ref{AngStabTraj2_rest}) in terms of the $\tau$ coordinate
defined in (\ref{Defr}) such that
\begin{align}\label{AngStabTraj3_z1}
z^1 = & -\epsilon \cosh \left( \frac{\tau}{2} \right) \, ,
\\\label{AngStabTraj3_z2}
z^2 = & \pm i \epsilon \sinh \left( \frac{\tau}{2} \right) \, ,
\\\label{AngStabTraj3_rest}
z^3 = & z^4 =  0 \, .
\end{align}
Comparing the above with the deformed conifold coordinates
(\ref{FullDefConifoldcoord1})-(\ref{FullDefConifoldcoord4}), while one needs to
solve transcendental equations in general, for sufficiently simple trajectory like
(\ref{AngStabTraj3_z1})-(\ref{AngStabTraj3_rest}), one can easily translate it in
terms of the restriction on angular coordinates
\begin{eqnarray}
\label{AngStabTraj4_1}
+&:&\theta_1=\theta_2=0\,,~~~\frac{\psi+(\phi_1+\phi_2)}{2}=\pi\,,
\\
\label{AngStabTraj4_2}
-&:&\theta_2=\theta_2=\pi\,,~~~\frac{\psi-(\phi_1+\phi_2)}{2}=0\,.
\end{eqnarray}
These two equivalent branches are the stabilized values of the isometry directions
broken by the Kuperstein embedding (\ref{Kuperemb1}). Exactly the same combinations
of angles also appear when one compares the angular stable trajectory in the
singular conifold (\ref{AngStaTra1}) with the corresponding embedding coordinates
written in terms of the angles. Since the embedding coordinates
(\ref{FullDefConifoldcoord1})-(\ref{FullDefConifoldcoord4}) interpolate the entire
throat, and
 that the stabilized angular values are the same in both asymptotic regions, it is
suggestive that  the angular stable trajectory
(\ref{AngStabTraj2_z1})-(\ref{AngStabTraj2_rest}) is valid not only in the 
{regions near or far away from the tip}
 but also for the entire deformed conifold. It would be interesting to
demonstrate this explicitly with the inflaton potential derived from the full warped
deformed conifold metric.

One should note that on each branch, the dependence on the combination
$[\psi\mp(\phi_1+\phi_2)]/2$ vanishes from the scalar potential
$V_F(z^1+\bar{z}^1,|z^1|^2,r,\sigma)$, despite the fact that $\psi$ and
$(\phi_1+\phi_2)/2$ appear individually in $z^1$ and $\bar{z}^1$. For each branch,
the corresponding combination can take arbitrary value {\em without} affecting the
resultant trajectory. Furthermore, it is also obvious that the combination
$(\phi_1-\phi_2)/2$ does not appear explicitly in $z^1$ or $\bar{z}^1$ hence in the
scalar potential. Both $(\phi_1-\phi_2)/2$ and one of $[\psi\mp(\phi_1+\phi_2)]/2$
are the residual isometries preserved by the stable inflationary trajectory in the
Kuperstein embedding (\ref{Kuperemb1}).

The presence of the light additional residual isometries can in principle give
significant contribution to the power spectrum by the Lyth effect we discussed in
Section~\ref{Lytheffect}, by coupling them with the canonical inflaton through the
tachyon potential at the end of the inflation. However the magnitude of such effect
is also controlled by the stabilized values of broken isometry directions, and the
dependence is encoded in the measure factors $\Gamma_2$ and $\Gamma_3$. We can
easily calculate them by using (\ref{Defgvier})-(\ref{eq:DeformMetric}) and writing
out explicitly the metrics of $S^2$ and $S^3$ in terms of the deformed conifold
angular coordinates as
\begin{align}\label{expS2S3}
d\Omega_2:&~\frac{(2/3)^{1/3}}{8}\left[(g_1)^2+(g_2)^2\right] \, ,
\\
d\Omega_3:&~\frac{(2/3)^{2/3}}{2}\left[(g_3)^2+(g_4)^2+\frac{1}{2}(g_5)^2\right] \,.
\end{align}
Restricting them to the specific trajectories (\ref{AngStabTraj4_1}) and
(\ref{AngStabTraj4_2}), we obtain
\begin{eqnarray}
+&:&~d\Omega_2=0\,,~~~d\Omega_3=\frac{(2/3)^{2/3}}{2} \left\{ d \left[ \psi +
(\phi_1+\phi_2) \right] \right\}^2=0 \, ,
\\
-&:&~d\Omega_2=0\,,~~~d\Omega_3=\frac{(2/3)^{2/3}}{2} \left\{ d \left[ \psi -
(\phi_1+\phi_2) \right] \right\}^2=0 \, .
\end{eqnarray}
Hence for both $(\phi_1-\phi_2)/2$ and $[\psi\mp(\phi_1+\phi_2)]/2$, their measure
factors $\Gamma_2$ and $\Gamma_3$ vanish identically along
(\ref{AngStabTraj2_z1})-(\ref{AngStabTraj2_rest}), or equivalently
(\ref{AngStabTraj4_1}) and (\ref{AngStabTraj4_2}). In other words, even though the
$D3$-$\overline{D3}$ angular separations $\Delta [(\phi_1-\phi_2)/2]$ and
$\Delta\{[\psi\mp(\phi_1+\phi_2)]/2\}$ can be finite, the proper separations along
these directions in fact vanish. Therefore, despite having the necessary conditions,
e.g. small slow-roll parameter $\varepsilon$ for the Lyth effect to be potentially
significant, the calculations here demonstrate that, due to the degeneracy of the
residual isometry directions, it in fact {\it does not} take place along the
specific angular stable trajectory considered. However, for other embeddings that
preserve some residual isometries on the $S^3$ at the tip of the conifold
\cite{D3vacua}, our
results in Section \ref{Lytheffect} can be used to
estimate the size of these end of inflation effects. One can easily use for example
the formula (\ref{Pend}) to obtain the ratio between the power spectrum at the
horizon exit and the end of inflation as
\begin{equation}\label{RatioPkPe}
\frac{\mathcal{P}_{\zeta_\fin}}{\mathcal{P}_k} =
\frac{\varepsilon_k}{\varepsilon_\fin} \frac{1}{2 \left[ (\vartheta_{\rm
c}/\vartheta_{\fin})^2-1 \right]} \, .
\end{equation}
The ratio
$\varepsilon_k/\varepsilon_\fin$ can be as large as $\cO(1)$\footnote{See
the discussion in Appendix~\ref{app_Baumann}.},
while $(\theta_c/\theta_\fin)^2 \gtrsim 1$,
therefore in the scenario we described eariler, where Coulombic attraction is
decoupled, $ \mathcal{P}_{\zeta_\fin}$ can possibly give comparable
contribution to
the power spectrum
$\mathcal{P}_k$.

\section{Discussion}
\label{Discussion}
\setcounter{equation}{0}

In this paper, we studied the systematics of multi-field effects at the end of
warped $D$ brane inflation. We discussed the necessary criteria for the isocurvature
perturbations generated by the angular motion of a mobile $D3$ brane to be converted
into the curvature perturbations usually associated with its radial motion in this
scenario. We found that the significance of the end of inflation effects considered
in Ref.~\cite{Lyth} depends on the specific mechanism for uplifting the vacuum
energy. If the uplifting is due to some distant $\overline{D3}$ branes or a $D$-term
potential, the Coulombic potential can easily become subdominant even towards the
end of inflation,
and the effects described in Ref.~\cite{Lyth} can in principle be significant.
However, in the most explicit $D$ brane
inflation constructed to date \cite{Baumann0,Baumann1}, the $D7$ brane embedding
chosen \cite{Kuper} does not yield such effects, { regardless of the uplifting
mechanism.} This latter result is specific to the embedding of the moduli
stabilizing branes as well as the infrared geometry of the throat. %
Along the stable trajectory for the embedding considered in
Ref.~\cite{Kuper}, the proper distance for the residual isometry direction vanishes
in the entire
throat, the moduli space vanishes at the tip. %
It would be interesting to
examine other $D7$ brane embeddings and/or other warped throats which leave a moduli
space of vacua at the tip. Examples of such embeddings for the deformed conifold
appeared in Ref.~\cite{D3vacua}, where the residual isometry directions reside on
the finite size $S^3$. However, finding an angular stable trajectory in these
examples may remain challenging. Nevertheless, our results underscore the importance
of multi-field effects in string inflation, as noted also in the context of DBI
inflation recently in Ref.~\cite{Langlois} (see also earlier discussions in
Refs.~\cite{Huang:2007hh,Easson:2007dh}).

As discussed in Section~\ref{Lytheffect},
{the strength of the Lyth effect}
depends on the ratio
$\varepsilon_k/\varepsilon_e$. Since the flat region of the inflaton potential
considered in Refs.~\cite{Baumann0,Baumann1} is an inflection point, $\varepsilon_k$
depends sensitively on where around the inflection point corresponds to the CMB
scale. Given a $D$ brane inflation model which can yield
{the Lyth effect}
considered here, a precise determination of the amplitude of such
effects would require the use of the full KS metric~\cite{KS}. {This is yet another
context in which details of the warped geometries in the infrared can have
significant effects on the CMB observations~\cite{inflation_tip}.}
{Furthermore, regardless of the
{Lyth effect}
 studied here,} a detailed
comparison of the WMAP data with microscopic parameters of $D$ brane inflation
requires identifying the relevant part of the inflaton potential which generates the
observed CMB
{anisotropy},
 and the full KS metric is essential. Work along these
lines is underway.

Finally, one
may hope to also realize the curvaton mechanism~\cite{curvaton} using these light fields.
In the setup {we discussed}, however, inflation ends as $D3$ and
$\overline{D3}$ annihilate and thus the would-be curvaton fields themselves
disappear. For the same reason, any multi-field effect~\cite{multi}
after inflation will not be present as long as they are associated with $D3$ or
$\overline{D3}$ branes. Nevertheless it would be interesting to implement the curvaton
scenario in a different setup satisfying a number of
constraints~\cite{Gong:2006hf}.

\subsection*{Acknowledgement}
\paragraph{}

We are indebted to Bret Underwood for numerous valuable insights and discussions,
and for collaboration at the initial stage of this project.
 We are also grateful to Misao Sasaki for discussions and comments on the manuscript.
We thank Daniel Baumann, Chong-Sun Chu, Min-Xin Huang, David Lyth, Fernando Marchesano,
 Liam McAllister,
Peter Ouyang, Sudhakar Panda, and Fernando Quevedo for helpful discussions.
 HYC would like to thank KITP at USCB and the Sixth Simons Workshop at SUNY Stony
Brook for their hospitalities where part of the work was being carried out. JG is
grateful to the Santa Fe 08 Cosmology Summer Workshop for hospitality where this
work was being finished.
 The work of HYC and GS is supported in part by NSF Career Award No. PHY-0348093, DOE
grant DE-FG-02-95ER40896, a Research Innovation Award and a Cottrell Scholar Award
from Research Corporation, and a Vilas Associate Award from the University of
Wisconsin.
 JG is partly supported by the Korea Research Foundation Grant KRF-2007-357-C00014
funded by the Korean Government.

\appendix

\section{Details of the warped deformed conifold}
\label{app_conifold}
\setcounter{equation}{0}
\paragraph{}

Here we collect a few facts concerning the various coordinates parameterizing the
deformed conifold.  It is defined via the equation
\begin{eqnarray}
\sum_{A=1}^4 (z^A)^2 = \epsilon^2 \,,
\end{eqnarray}
and the $D7$ brane embeddings we use are given in terms of one or the other of these
sets of coordinates.  These coordinates can be related to coordinates on the $S^3$
at the bottom of the throat as follows. We follow Ref.~\cite{Candelas} with some
modifications to their notation. We define the matrix $W$ as
\begin{equation}
W \equiv L W_0 R^\dagger \, ,
\end{equation}
with
\begin{equation}
W_0 \equiv
\left(%
\begin{array}{cc}
  \epsilon/\sqrt{2} & \sqrt{r^3 - \epsilon^2} \\
  0 & -\epsilon/2 \\
\end{array}%
\right) \, ,
\end{equation}
where $L$ and $R$ are $SU(2)$ matrices parameterized by three Euler angles (We are
using the standard $r$-variable on the conifold, related to that in
Ref.~\cite{Candelas} by $r = r_\mathrm{there}^{2/3}$). We choose the convention
\begin{eqnarray}
\label{Wdef} W = \begin{pmatrix}
-w_3 \; w_2 \cr -w_1 \; w_4
\end{pmatrix}
= - \frac{1}{\sqrt{2}}
\begin{pmatrix}
z^3 + i z^4 \;\;\;\; z^1 - i z^2 \cr z^1 + i
z^2 \;\; -z^3 + i z^4
\end{pmatrix} \,,
\end{eqnarray}
where we have chosen the $w$'s so as to agree with (32)-(35) of Ref.~\cite{Gaugino}
when we use the parameterization of Euler angles given in (2.24) and (2.25) of
Ref.~\cite{Candelas}.  One indeed finds that
\begin{eqnarray}
{\rm det}\ W = w_1 w_2 - w_3 w_4 = -\frac{1}{2} \sum_{A=1}^4
(z^A)^2 =  - \frac{1}{2} \epsilon^2 \,,\label{Defwcoord}
\end{eqnarray}
as required. At generic $r > \epsilon^{2/3}$, one of the six Euler angles in $L$
and $R$ is redundant, and the remaining five along with $r$ parameterize the
deformed conifold.  For $ r \gg \epsilon^{2/3}$ the deformed conifold is well
approximated by the singular conifold, with the angles parameterizing $T^{1,1}$.

The complex embedding coordinates of deformed conifold $\{z^1,z^2,z^3,z^4\}$ can be
expressed in terms of the real coordinates
$\{\tau\in{\mathbb{R}\,,~\psi\in[0,4\pi]\,,~\theta_{1,2}\in[0,\pi]\,,~\phi_{1,2}\in[0,2\pi]}\}$,
$\Xi=\tau+i\psi$ as
\begin{align}
\label{FullDefConifoldcoord1}
z^1 = & \epsilon \left[ \cosh \left(\frac{\Xi}{2}\right) \cos
\left(\frac{\theta_1+\theta_2}{2}\right) \cos\left(\frac{\phi_1+\phi_2}{2}\right) +
i \sinh \left(\frac{\Xi}{2}\right) \cos \left(\frac{\theta_1-\theta_2}{2}\right)
\sin \left(\frac{\phi_1+\phi_2}{2}\right) \right] \, ,
\\
\label{FullDefConifoldcoord2}
z^2 = & \epsilon \left[ -\cosh \left(\frac{\Xi}{2}\right) \cos
\left(\frac{\theta_1+\theta_2}{2}\right) \sin \left(\frac{\phi_1+\phi_2}{2}\right) +
i \sinh \left(\frac{\Xi}{2}\right) \cos \left(\frac{\theta_1-\theta_2}{2}\right)
\cos \left(\frac{\phi_1+\phi_2}{2}\right)\right] \, ,
\\
\label{FullDefConifoldcoord3}
z^3 = & \epsilon \left[ -\cosh \left(\frac{\Xi}{2}\right) \sin
\left(\frac{\theta_1+\theta_2}{2}\right) \cos \left(\frac{\phi_1-\phi_2}{2}\right) +
i \sinh \left(\frac{\Xi}{2}\right) \sin \left(\frac{\theta_1-\theta_2}{2}\right)
\sin \left(\frac{\phi_1-\phi_2}{2}\right)\right] \, ,
\\
\label{FullDefConifoldcoord4}
z^4 = & \epsilon \left[ -\cosh \left(\frac{\Xi}{2}\right) \sin
\left(\frac{\theta_1+\theta_2}{2}\right) \sin \left(\frac{\phi_1-\phi_2}{2}\right) -
i \sinh \left(\frac{\Xi}{2}\right) \sin \left(\frac{\theta_1-\theta_2}{2}\right)
\cos\left(\frac{\phi_1-\phi_2}{2}\right)\right] \, .
\end{align}
At the tip of the throat $r=\epsilon^{2/3}$, we can reduce the complex coordinates
$z^A$ in terms of the angles of the $S^3$ $\{\theta\,,\omega\,,\phi\}$
as\footnote{Note that the exact relation between these coordinates and those of
(\ref{FullDefConifoldcoord1})-(\ref{FullDefConifoldcoord4}) can be obtained by
identifying the non-vanishing $S^3$ in the metric using the vielbeins defined in
the next subsection.}
\begin{align}\label{z3Coords}
z^1 = & \epsilon \sin \left( \frac{\theta}{2} \right) \sin \left( \frac{\omega -
\phi}{2} \right) \, ,
\\
z^2 = & \epsilon \sin \left( \frac{\theta}{2} \right) \cos \left( \frac{\omega -
\phi}{2} \right) \, ,
\\
z^3 = & \epsilon \cos \left( \frac{\theta}{2} \right) \cos \left( \frac{\omega +
\phi}{2} \right) \, ,
\\
z^4 = & \epsilon \cos \left( \frac{\theta}{2} \right) \sin \left( \frac{\omega +
\phi}{2} \right) \, .
\end{align}
We see that in this case, $S^3$ is a real slice of each $z^\al$ coordinate and the
metric is given by
\begin{equation}\label{S3metric}
d\Omega_{3} = (d\omega + \cos\theta d\phi)^2 + d\theta^2 + \sin^2\theta d\phi^2 \, .
\end{equation}

\subsection{Metric}
\paragraph{}

It is convenient to work in a diagonal basis of the metric by using the basis of one
forms~\cite{KS}
\begin{align}\label{Defgvier}
g^1 \equiv & \frac{e^1-e^3}{\sqrt{2}} \, ,~~~
g^2 \equiv  \frac{e^2-e^4}{\sqrt{2}} \, ,\nn
\\
g^3 \equiv & \frac{e^1+e^3}{\sqrt{2}} \, ,~~~
g^4 \equiv \frac{e^2+e^4}{\sqrt{2}} \, ,\nn
\\
g^5 \equiv & e^5 \, ,
\end{align}
where
\begin{align}\label{Defevier}
e^1 \equiv & -\sin\theta_1 d\phi_1 \, ,
\\
e^2 \equiv & d\theta_1 \, ,
\\
e^3 \equiv & \cos \psi \sin \theta_2 d\phi_2 - \sin\psi d\theta_2 \, ,
\\
e^4 \equiv & \sin\psi \sin \theta_2 d\phi_2 + \cos \psi d\theta_2 \, ,
\\
e^5 \equiv & d\psi + \cos \theta_1 d\phi_1 + \cos\theta_2 d\phi_2 \, .
\end{align}
The metric of the deformed conifold is then
\begin{equation}
ds_6^2 = \frac{1}{2} \epsilon^{4/3} K(\tau) \left\{ \frac{1}{3 [K(\tau)]^3} [d\tau^2
+(g^5)^2] + \cosh^2 \left( \frac{\tau}{2} \right) \left[ (g^3)^2 + (g^4)^2 \right] +
\sinh^2 \left( \frac{\tau}{2} \right) \left[(g^1)^2 + (g^2)^2 \right] \right\} \, ,
\label{eq:DeformMetric}
\end{equation}
where
\begin{equation}\label{DefKtau}
K(\tau) = \frac{\left[ \sinh(2\tau)-2\tau \right]^{1/3}}{2^{1/3}\sinh\tau} \, .
\end{equation}
The ten dimensional metric takes the warped form
\begin{equation}\label{Def10Dwarpmetric}
ds_{10}^2 = e^{2A(y)} \eta_{\mu\nu} dx^\mu dx^\nu + e^{-2A(y)} ds_6^2 \, ,
\end{equation}
where the warp factor is given by the expression~\cite{KS}
\begin{equation}
e^{4A(\tau)} = 2^{2/3} (g_s M \alpha')^2 \epsilon^{-8/3} I(\tau) \, ,
\end{equation}
where
\begin{equation}
I(\tau) \equiv \int_{\tau}^\infty dx\ \frac{x \coth x -1}{\sinh^2x} \left[ \sinh(2x)
- 2x \right]^{1/3} \, .
\end{equation}

\subsection{Little K\"ahler potential}
\paragraph{}

The warped deformed conifold metric (\ref{eq:DeformMetric}) can be obtained from the
``little'' K\"ahler potential $k(z^\alpha,\bar{z}^{\bar{\beta}})$ as
\begin{equation}
\tilde{g}_{\alpha\bar{\beta}} = \partial_\alpha \partial_{\bar{\beta}} k\, .
\end{equation}
Because the angular directions of the warped deformed conifold are isometries they
do not appear explicitly in the little K\"ahler potential, and in general the
K\"ahler potential only depends on the radial coordinate $\tau$
through~\cite{Candelas}
\begin{equation}
k(\tau) = \frac{\epsilon^{4/3}}{2^{1/3}} \int_0^\tau d\tau' \left[
\sinh(2\tau')-2\tau' \right]^{1/3} \, ,
\label{DefKahdefcon}
\end{equation}
where without loss of generality we set the integration constant to zero.  Using the
relation between $\tau$ and $r$, we can approximately solve for the large and small
$r$ limits as
\begin{equation}\label{asymptotic_k}
k(r) \to \left\{
\begin{split}
& \frac{3}{2}r^2 & \mbox{for } & r \gg \epsilon^{2/3} \, ,
\\
& k_0 + \frac{c}{\epsilon^{2/3}}(r^3 - \epsilon^2) & \mbox{for } & r \approx
\epsilon^{2/3} \, ,
\end{split}
\right.
\end{equation}
where $c = 2^{1/6}/3^{1/3} \approx 1.61887$.

\section{Brief review of the ``Delicate Universe''}
\setcounter{equation}{0}
\label{app_Baumann}
\paragraph{}

In Refs.~\cite{Baumann0, Baumann1}, the authors considered the region of large
$D3$-$\overline{D3}$ separation, so that the deformed conifold can be approximated
by its singular limit. The expression for the $F$-term potential (\ref{DefVF}) is
then greatly simplified. The non-perturbative superpotential is generated by $D3$ or
$D7$ brane wrapping a four cycle of the conifold (made compact by the bulk
geometry). Further, their presence partially breaks the full $SO(4)$ isometry group
of the deformed conifold. For example, consider the simplest Kuperstein embedding
given by holomorphic function (\ref{Kuperemb1}) which breaks $SO(4)$ down to an
$SO(3)$ subgroup rotating $\{z^2,z^3,z^4\}$. The trajectory of the canonical
inflaton then further breaks it to $SO(2)$. One should note here that in the
presence of the bulk NS-NS B-field, the $D7$ brane embedding (\ref{Kuperemb1}) can
remain supersymmetric without additional worldvolume flux: by contrast, the
supersymmetric $D7$ embeddings considered in the singular conifold limit as given in
Refs.~\cite{Ouyang, KarchKatz, ACR} can only remain supersymmetric on these four
cycles in the deformed conifold with additional worldvolume flux turned
on~\cite{COS1}.

Using the singular conifold metric and (\ref{Kuperemb1}) to calculate $V_F$, the
authors then included the Coulombic potential $V_{D3\overline{D3}}$ as given by
(\ref{DefVDuplifting}) such that the cancellation of the negative vacuum energy of
$V_F$ is due to the combination of the $\overline{D3}$ branes at the tip of the
deformed conifold and distant bulk. They stabilize the isometry directions broken by
$D7$ branes, and the resulting angular stable trajectory is given by
\begin{equation}\label{AngStaTra1}
z^1=-\frac{r^{3/2}}{\sqrt{2}} \leftrightarrow
\frac{\partial(V_F+V_{D3\overline{D3}})}{\partial \Psi_i}=0\,.
\end{equation}
Here $\{\Psi_i\}$ runs through the broken isometry directions and (\ref{AngStaTra1})
also imposes constraints on other embedding coordinates $z^2=\pm i
z^1\,,~z^3=z^4=0$. The axion $\varsigma$ can also be stabilized by tuning
{the
perturbative superpotential $W_0$ to be negative}. To proceed obtaining single field
inflation, an adiabatic approximation is taken to stabilize the volume modulus
$\sigma$ by solving the equation
\begin{equation}\label{volumestable}
\left.\frac{\partial(V_F+V_{D3\overline{D3}})(r,\sigma)}{\partial
\sigma}\right|_{\sigma_\ast}=0 \, .
\end{equation}
The authors approximated the solution of (\ref{volumestable}) by
\begin{equation}
\sigma_\ast(r)\approx\sigma_0\left[1+c_{3/2}\frac{r^2}{(2\mu^2)^{2/3}}\right]\,,
\end{equation}
where $\mu$ is the embedding parameter in (\ref{Kuperemb1}) and the
coefficient $c_{3/2} \approx [1 - 1/(2a\sigma_F)]/(n\sigma_F)$, and $\sigma_0$ is
the stabilized volume at the tip of the throat after including the uplifting term
$V_{D3\overline{D3}}$. Finally, the canonical inflaton $\phi$ was related to the
radial coordinate of the mobile $D3$ brane via (\ref{Inflatonr}).

\begin{figure}[h]
\begin{center}
\includegraphics[width = 8cm]{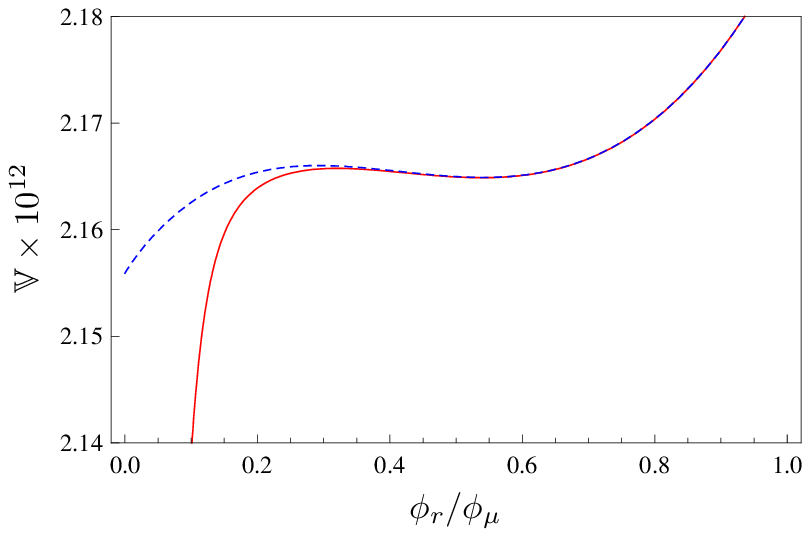}
\includegraphics[width = 8cm]{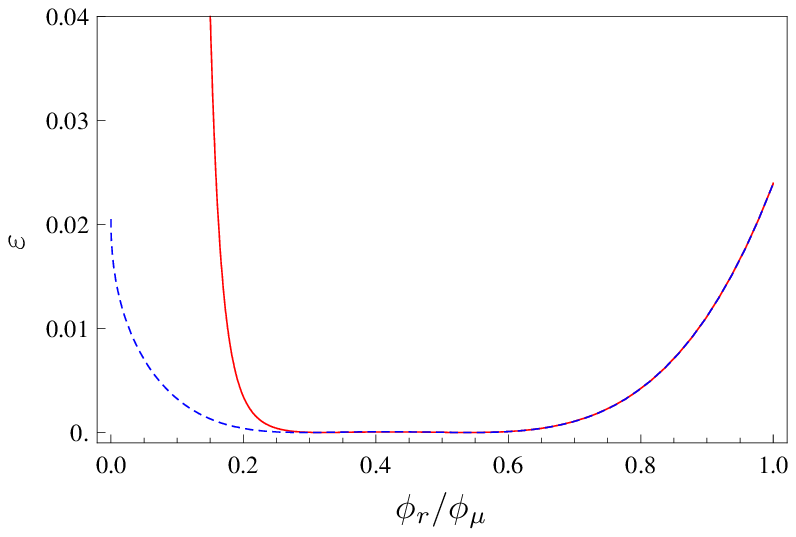}
\end{center}
\begin{center}
\includegraphics[width = 8cm]{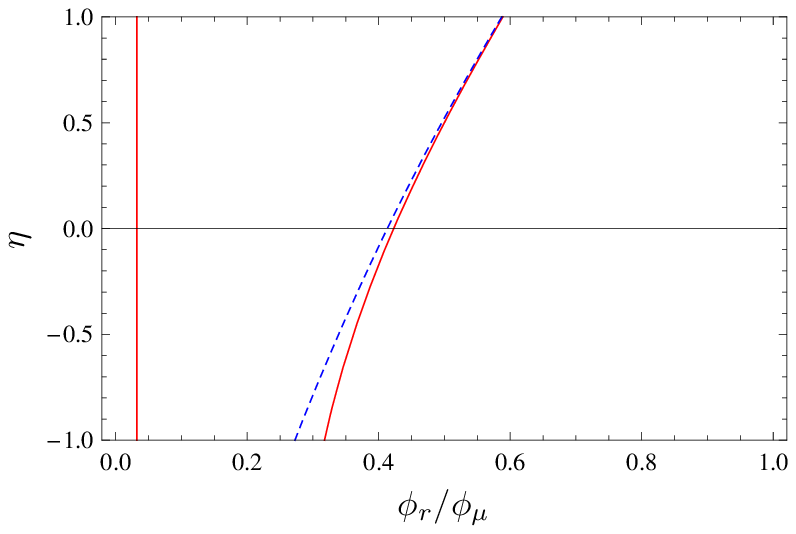}
\includegraphics[width = 8cm]{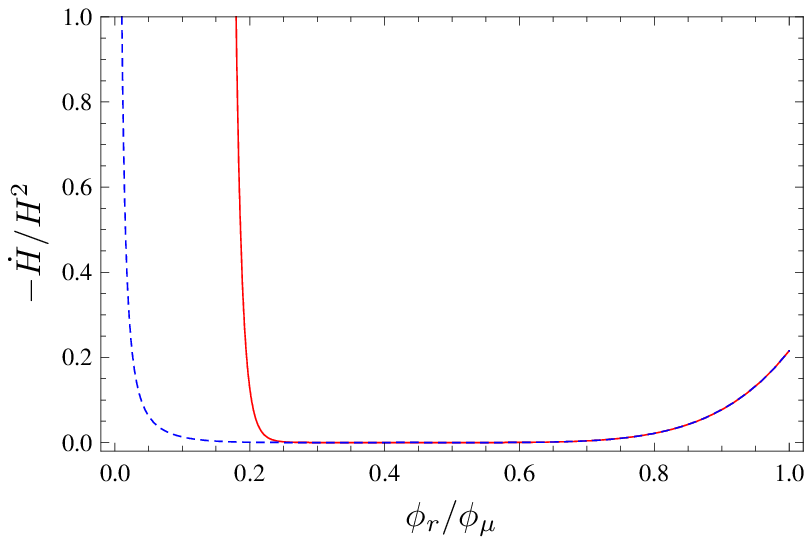}
\end{center}
\caption{(Upper left) the inflaton potential (\ref{delicatepotential}) and the
resulting slow-roll parameters, (upper right) $\varepsilon$ and (lower left) $\eta$.
We show both cases where the Coulombic piece proportional to $r^{-4}$ is present
(solid line) and absent (dotted line). As can be seen, without the Coulombic term
inflation proceeds deep inside the throat, i.e. very small $r$ region, but
(\ref{delicatepotential}) is no more valid there. In the lower right panel, we show
$-\dot{H}/H^2$ which is {\em exactly} equivalent to the acceleration of the scale
factor: $-\dot{H}/H^2 < 1$ means acceleration. Clearly, the criteria $|\eta| = 1$
{\em does not} guarantee that inflation ends at the corresponding point, especially
when the Coulombic term is negligible which is the scenario we discussed in the main
text.}
 \label{DUplot}
\end{figure}

Putting everything together, the single inflaton potential derived at large brane
separation for the trajectory (\ref{AngStaTra1}) is subsequently given by
\begin{align}\label{delicatepotential}
{\mathbb{V}}(\phir) = & \frac{\kappa^2
a|A_0|^2e^{-2a\sigma_\ast}}{3[U(\phir,\ss)]^2}h(\phir)^{2/n} \left\{ 2a\sigma_\ast + 6 -
6e^{a\sigma_\ast}\frac{|W_0|}{|A_0|}h(\phir)^{1/n} + \frac{3}{n h(\phir)}\frac{\phir}{\phim}
\left[ c_0 - h(\phir)\sqrt{\frac{\phir}{\phim}} \right] \right\}
\nn\\
& + \frac{D_0+D_{\rm others}}{[U(\phir,\sigma_\ast)]^2} \, ,
\end{align}
where
\begin{align}\label{BaumannDefinitons}
h(\phir) = & 1 + \left( \frac{\phir}{\phim} \right)^{3/2} \, ,
\\
c_0 = & \frac{9}{4na\sigma_0\phi_{\mu}^2/\mpl^2} \, ,
\\
\phim^2 = & \frac{3}{2}T_3(2\mu^2)^{2/3} \, .
\end{align}
Here the approximation $D(\phir) \approx D_0+D_{\rm others}$ is taken for large
radius $r\gg\epsilon^{2/3}$. The explicit inflaton potential
(\ref{delicatepotential}) represents one of the most well developed and top-down
brane inflation model to date, which explicitly includes the effects of
compactification and moduli stabilization. To obtain sufficiently flat region of
${\mathbb{V}}(\phi)$, the parameters in the inflaton potential need to be
`delicately' tuned. For the specific set of parameters considered in
Ref.~\cite{Baumann1} (see below Fig.~2 there), the inflaton potential
${\mathbb{V}}(\phi)$ has a sharp drop, however it is induced by the
$D3$-$\overline{D3}$ Coulombic interaction which only becomes significant near the
tip of the throat, without which the inflaton potential is in fact smooth and
inflation continues until much smaller radius into the deformed conifold\footnote{We
thank Daniel Baumann for communicating about this issue.}. However in such region,
the singular conifold approximation should break down. In Fig.~\ref{DUplot} we show
the effective potential (\ref{delicatepotential}) and the related slow-roll
parameters\footnote{To plot $-\dot{H}/H^2$, we have used~\cite{Gong:2001he}

\begin{align}
-\frac{\dot{H}}{H^2} = & \frac{\mpl^2}{2} \left(
\frac{\mathbb{V}_{,\phir}}{\mathbb{V}} \right)^2 - \frac{\mpl^4}{3} \left(
\frac{\mathbb{V}_{,\phir}}{\mathbb{V}} \right)^4 + \frac{\mpl^4}{3} \left(
\frac{\mathbb{V}_{,\phir}}{\mathbb{V}} \right)^2
\frac{\mathbb{V}_{,\phir\phir}}{\mathbb{V}}
\nonumber\\
& + \frac{4}{9}\mpl^6 \left( \frac{\mathbb{V}_{,\phir}}{\mathbb{V}} \right)^6 -
\frac{5}{6} \mpl^6 \left( \frac{\mathbb{V}_{,\phir}}{\mathbb{V}} \right)^4
\frac{\mathbb{V}_{,\phir\phir}}{\mathbb{V}} + \frac{5}{18} \mpl^6 \left(
\frac{\mathbb{V}_{,\phir}}{\mathbb{V}} \right)^2 \left(
\frac{\mathbb{V}_{,\phir\phir}}{\mathbb{V}} \right)^2 + \frac{\mpl^6}{9} \left(
\frac{\mathbb{V}_{,\phir}}{\mathbb{V}} \right)^2
\frac{\mathbb{V}_{,\phir}\mathbb{V}_{,\phir\phir\phir}}{\mathbb{V}^2} + \cdots \, .
\end{align}
}.

Of course, it is possible that one can try to select a different set of parameters
such that the slow-roll parameters $\varepsilon$ and $\eta$ become order one at much
larger radius without including the Coulombic interaction. The point we would like
to emphasize is that for the purpose of parameter scanning, consistently excluding
the tip region in the analysis of inflation imposes further constraints (in addition
to obtaining a sufficient number of $e$-folds, and the correct amplitude of the
power spectrum, etc). However if we relax such constraints and allow the
inflationary epoch to extend deep into the deformed conifold region, one needs to
take into account the full deformed conifold metric. As we have explicitly shown in
the main text, $\varepsilon$ can remain small in this region, using the metric near
the tip of deformed conifold. In other words, inflation ends when the canonical
inflaton reaches its limit in field range, rather than when $\varepsilon$ becomes
large. In such case, there can potentially be an additional contribution to the
curvature perturbation arising from quantum fluctuations in the light residual
isometry directions, which can significantly modify the estimates made far from the
tip.

In relation to the scenario we proposed in the main text, where $V_{\rm Coulomb}$ is
neglected, the potential (\ref{delicatepotential}) should be regarded as the
ultraviolet completion of our inflaton potential (\ref{MathbbVphi}) with uplifting
exclusively done by distant $\overline{D3}$ branes. Assuming that the flat region in
(\ref{delicatepotential}) (near its inflection point) corresponds to the large
observable scales, and that most but not all the $e$-folds are generated there, this
allows us to have an estimate of $\varepsilon_k$ near the horizon exit. To make such
statement precise of course requires the calculation of the inflaton potential with
respect to the full deformed conifold metric, this is an interesting although
potentially challenging direction, which we shall return in the near future.

\section{Stability analysis for angular extremum trajectory}
\label{app_stability}
\setcounter{equation}{0}
\paragraph{}

In this appendix we will explicitly obtain the angular extremum trajectory for the
Kuperstein embedding~(\ref{Kuperemb1}) in the near tip region of deformed conifold,
and demonstrate its stability.

First, without lost of generality, we take both $\mu$ and $\epsilon$ to be real,
and let us write the $F$-term scalar potential $V_F=V_\mathrm{KKLT}+\Delta V_F$ in
the following form:
\begin{align}\label{VKKLT}
V_\mathrm{KKLT} = & \frac{2a\kappa^2|A_0|^2e^{-2a\sigma}}{[U(r,\sigma)]^2} \left| 1 -
\frac{z_1}{\mu} \right|^{2/n} \left\{ 1 - e^{a\sigma}\frac{|W_0|}{|A_0|} \left| 1 -
\frac{z_1}{\mu} \right|^{-1/n} + \frac{a}{6} \left[ 2\sigma - \gamma k_0 + \gamma c
\epsilon^{4/3} \left( 1 - \frac{{\epsilon^2}}{r^3} \right) \right] \right\}
\nonumber\\
= & \mathcal{A}(r,\sigma) \left( 1 - \frac{z_1+\bar{z}_1}{\mu} +
\frac{|z_1|^2}{\mu^2} \right)^{1/n} \left\{ 1 - e^{a\sigma}\frac{|W_0|}{|A_0|}
\left( 1 - \frac{z_1+\bar{z}_1}{\mu} + \frac{|z_1|^2}{\mu^2} \right)^{-1/(2n)} +
\mathcal{B}(r,\sigma) \right\} \, ,
\end{align}
and
\begin{align}\label{DeltaVF}
\Delta{V}_F = & \frac{\kappa^2|A_0|^2e^{-2a\sigma}}{3n^2[U(r,\sigma)]^2} \left| 1 -
\frac{z_1}{\mu} \right|^{2(1/n-1)} \left\{ \frac{\epsilon^{2/3}}{c\mu^2\gamma}
\left( 1 - \frac{|z_1|^2}{r^3} \right) + \frac{an}{\mu} \left[ \left( 1 -
\frac{\bar{z}_1}{\mu} \right) \left(z_1 - \bar{z}_1 \frac{\epsilon^2}{r^3}
\right) + c.c. \right] \right\}
\nonumber\\
= & \mathcal{C}(r,\sigma) \left( 1 - \frac{z_1+\bar{z}_1}{\mu} +
\frac{|z_1|^2}{\mu^2} \right)^{1/n-1}
\nonumber\\
& \times \left\{ \frac{\epsilon^{2/3}}{c\mu^2\gamma} \left( 1 -
\frac{|z_1|^2}{r^3} \right) + an \left[ \frac{z_1+\bar{z}_1}{\mu} \left( 1 -
\frac{\epsilon^2}{r^3} \right) + \frac{(z_1+\bar{z}_1)^2}{\mu^2}
\frac{\epsilon^2}{r^3} - \frac{2|z_1|^2}{\mu^2} \left(1 +
\frac{\epsilon^2}{r^3} \right) \right] \right\} \, .
\end{align}
Such explicit forms (\ref{VKKLT}) and (\ref{DeltaVF}) will be useful in the
subsequent stability analysis. Note that both $\mathcal{A}$ and $\mathcal{C}$ have
mass dimension 4 and the remaining terms are dimensionless, and we have rewritten
the expressions in terms of $|z_1|^2$ and $z_1+\bar{z}_1$ wherever possible. From
(\ref{VKKLT}) and (\ref{DeltaVF}), $V_F$ now becomes a function of $\sigma$, $r$,
$z_1+\bar{z}_1$ and $|z_1|^2$. To extract the light degree of freedom among all the
isometry directions, we first try to stabilize as many angular directions explicitly
broken by the presence of $D7$ as possible.

Recalling the analysis of Ref.~\cite{Baumann1}, where the trajectory in the singular
conifold along which the linear variations of $|z_1|^2$ and $z_1+\bar{z}_1$ vanish,
we can again apply this analysis and write down the variation of $z_1$ being
\begin{equation}\label{variationz1}
\delta z_1^{(0)}=\sum_{j=2}^{4}\alpha_j z^{(0)}_j \ ,
\end{equation}
with $\alpha_i\in{\mathbb{R}}$. Here $\left\{
z^{(0)}_1,z^{(0)}_2,z^{(0)}_3,z^{(0)}_4 \right\}$ are the coordinates of a fiducial
point and from here
${\alpha_2,\alpha_3,\alpha_4,\beta_3,\beta_4}\subset \{\Psi_i\}$ are local
coordinates on the base of the cone.
Vanishing of the linear variations can then be written as
\begin{align}
\delta|z_1|^2 = & \sum^4_{j=2}\alpha_j \left[
z^{(0)}_1\bar{z}^{(0)}_j+\bar{z}^{(0)}_1z^{(0)}_j \right] = 0 \, ,
\label{variation1}\\
\delta(z_1+\bar{z}_1) = & i\sum_{j=2}^4 \alpha_j \left[ z^{(0)}_j-\bar{z}_j^{(0)}
\right] = 0 \, .
\label{variation2}
\end{align}
For (\ref{variation1}) to be satisfied for all $\{\alpha_i\}$, we need to have
\begin{equation}
z_j^{(0)}=i\varrho_j z_1^{(0)} \, ,
\end{equation}
where $\varrho_j \in \mathbb{R}$. Using the $SO(3)$ symmetry, one can set
$\varrho_2\neq 0$ while $\varrho_3=\varrho_4=0$. (\ref{variation2}) then implies
$z_1^{(0)}$ is strictly real. Subjecting $z^{(0)}_1$ and $z^{(0)}_2$ to the
constraint $\left( z_1^{(0)} \right)^2 + \left( z_2^{(0)} \right)^2 = \epsilon^2$
and the definition $r^3 = |z_1|^2 + |z_2|^2 = \left( z_1^{(0)} \right)^2 + \left|
z_2^{(0)} \right|^2$, we can see that
\begin{equation}
\varrho_2 = \pm \sqrt{\frac{r^3 - \epsilon^2}{r^3 + \epsilon^2}} \, ,
\end{equation}
leading to the angular extremum trajectory along
\begin{align}
\label{AngStabtraj_z1}
z_1^{(0)} = & \pm\sqrt{\frac{r^3+\epsilon^2}{2}} \, ,
\\
\label{AngStabtraj_z2}
z_2^{(0)} = & \pm i \sqrt{\frac{r^3-\epsilon^2}{2}} \, .
\end{align}
Notice that in the singular conifold limit $\epsilon\to 0$, (\ref{AngStabtraj_z1})
and (\ref{AngStabtraj_z2}) reduce to the one in (\ref{AngStaTra1}).

Let us now proceed with the stability analysis for (\ref{AngStabtraj_z1}) and
(\ref{AngStabtraj_z2}). We first notice that along these trajectories the linear
perturbations in $|z_1|^2$ and $z_1+\bar{z}_1$ disappear, and we can further see
that
\begin{align}
z_1 = & z_1^{(0)} \left[ 1 - \frac{1}{2} \left( \alpha_2^2 + \alpha_3^2 + \alpha_3^2
\right) + \frac{i}{2}\varrho_2 \left( 2\alpha_2 - \alpha_3\beta_3 - \alpha_4\beta_4
\right) + \cdots \right] \, ,
\\
z_1 + \bar{z}_1 = & 2z_1^{(0)} \left[ 1 - \frac{1}{2} \left( \alpha_2^2 + \alpha_3^2
+ \alpha_3^2 \right) + \cdots \right] \, ,
\\
|z_1|^2 = & {z_1^{(0)}}^2 \left[ 1 - \left( \frac{2\epsilon^2}{r^3 +
\epsilon^2}\alpha_2^2 + \alpha_3^2 + \alpha_4^2 \right) + \cdots \right] \, .
\end{align}
Then we find that
\begin{align}
\left.\frac{\partial^2|z_1|^2}{\partial\Psi_i\partial\Psi_j} \right|_0 = & -(r^3 +
\epsilon^2) \left( \frac{2\epsilon^2}{r^3 +
\epsilon^2}\delta_{i2}\delta_{j2} + \delta_{i3}\delta_{j3} +
\delta_{i4}\delta_{j4} \right) \, ,
\\
\left.\frac{\partial^2(z_1+\bar{z}_1)}{\partial\Psi_i\partial\Psi_j} \right|_0 = &
\mp\sqrt{2(r^3 + \epsilon^2)} ( \delta_{i2}\delta_{j2} + \delta_{i3}\delta_{j3} +
\delta_{i4}\delta_{j4} ) \, ,
\end{align}
so that the mass matrix along the extremal trajectory is given by
\begin{equation}\label{massmatrix}
\left.\frac{\partial^2V}{\partial\Psi_i\partial\Psi_j}\right|_0 =
\left(%
\begin{array}{ccccc}
  X + 2\epsilon^2Y/(r^3+\epsilon^2) & 0 & 0 & 0 & 0 \\
  0 & X+Y & 0 & 0 & 0 \\
  0 & 0 & X+Y & 0 & 0 \\
  0 & 0 & 0 & 0 & 0 \\
  0 & 0 & 0 & 0 & 0 \\
\end{array}%
\right) \, ,
\end{equation}
where
\begin{align}
X = & \mp\sqrt{2(r^3 + \epsilon^2)} \left.
\frac{\partial{V}}{\partial(z_1+\bar{z}_1)} \right|_0 \, ,
\\
Y = & -(r^3 + \epsilon^2) \left. \frac{\partial{V}}{\partial|z_1|^2} \right|_0\,.
\end{align}
Therefore, three angular directions, viz. $\alpha_2$, $\alpha_3$ and $\alpha_4$ have
definite masses squared no matter positive or negative, while $\beta_3$ and
$\beta_4$ remain perfectly flat unless any other effect which breaks the symmetry is
introduced, e.g. bulk mass terms. Hence, if we are to look for light angular
directions which can give rise to interesting and/or dangerous effects at the end of
inflation, there are two of them provided that the other three directions are
stabilized. First we must check whether this is indeed achieved.

From (\ref{VKKLT}) and (\ref{DeltaVF}), after some calculations, we can find that
for $V_\mathrm{KKLT}$
\begin{align}
\left. \frac{\partial{V_\mathrm{KKLT}}}{\partial(z_1+\bar{z}_1)} \right|_0 = &
-\frac{\mathcal{A}}{n\mu} \left| 1 \mp \frac{\sqrt{r^3+\epsilon^2}}{\sqrt{2}\mu}
\right|^{-2(1 - 1/n)} \left( 1 - \frac{e^{a\sigma}}{2}\frac{|W_0|}{|A_0|} \left| 1
\mp \frac{\sqrt{r^3+\epsilon^2}}{\sqrt{2}\mu} \right|^{-1/n} + \mathcal{B}
\right)\,,
\\
\left. \frac{\partial{V_\mathrm{KKLT}}}{\partial|z_1|^2} \right|_0 = &
\frac{\mathcal{A}}{n\mu^2} \left| 1 \mp \frac{\sqrt{r^3+\epsilon^2}}{\sqrt{2}\mu}
\right|^{-2(1-1/n)} \left( 1 - \frac{e^{a\sigma}}{2}\frac{|W_0|}{|A_0|} \left| 1 \mp
\frac{\sqrt{r^3+\epsilon^2}}{\sqrt{2}\mu} \right|^{-1/n} + \mathcal{B} \right)\,,
\end{align}
and for $\Delta{V}_F$
\begin{align}
\left. \frac{\partial\Delta{V}_F}{\partial(z_1+\bar{z}_1)} \right|_0 = &
\frac{\mathcal{C}}{\mu} \left| 1 \mp \frac{\sqrt{r^3+\epsilon^2}}{\sqrt{2}\mu}
\right|^{-2(2-1/n)}
\nonumber\\
&\times \left\{ \left(1-\frac{1}{n}\right) \left[
\frac{\epsilon^{2/3}}{c\mu^2\gamma} \frac{r^3-\epsilon^2}{2r^3} - an \left( 1
- \frac{\varepsilon^2}{r^3} \right) \frac{\sqrt{2(r^3+\varepsilon^2)}}{\mu} \left(
\frac{\sqrt{r^3+\epsilon^2}}{\sqrt{2}\mu} \mp 1 \right) \right] \right.
\nonumber\\
& \hspace{0.5cm} \left. + an \left| 1 \mp
\frac{\sqrt{r^3+\epsilon^2}}{\sqrt{2}\mu} \right|^2 \left[ \left( 1 -
\frac{\epsilon^2}{r^3} \right) \pm \frac{2\sqrt{2(r^3+\epsilon^2)}}{\mu}
\frac{\epsilon^2}{r^3} \right] \right\} \, ,
\\
\left. \frac{\partial\Delta{V}_F}{\partial|z_1|^2} \right|_0 = &
-\frac{\mathcal{C}}{\mu^2} \left| 1 \mp \frac{\sqrt{r^3+\epsilon^2}}{\sqrt{2}\mu}
\right|^{-2(2-1/n)}
\nonumber\\
& \times \left\{ \left(1-\frac{1}{n}\right) \left[
\frac{\epsilon^{2/3}}{c\mu^2\gamma} \frac{r^3-\epsilon^2}{2r^3} - an \left( 1
- \frac{\epsilon^2}{r^3} \right) \frac{\sqrt{2(r^3+\epsilon^2)}}{\mu} \left(
\frac{\sqrt{r^3+\epsilon^2}}{\sqrt{2}\mu} \mp 1 \right) \right] \right.
\nonumber\\
& \left. \hspace{0.5cm} + \left| 1 \mp \frac{\sqrt{r^3+\epsilon^2}}{\sqrt{2}\mu}
\right|^2 \left[ \frac{\epsilon^{2/3}}{cr^3\gamma} + 2an \left(1 +
\frac{\epsilon^2}{r^3} \right) \right] \right\} \, .
\end{align}
Thus, from
\begin{equation}
\mathcal{C} = \frac{\mathcal{A}}{6an^2} \, ,
\end{equation}
we can write
\begin{align}
\label{X}
X = & \pm \frac{\mathcal{A}}{n} \frac{\sqrt{2(r^3+\epsilon^2)}}{\mu} \left| 1 \mp
\frac{\sqrt{r^3+\epsilon^2}}{\sqrt{2}\mu} \right|^{-2(1-1/n)}
\nonumber\\
& \times \left\{ 1 - \frac{e^{a\sigma}}{2}\frac{|W_0|}{|A_0|} \left| 1 \mp
\frac{\sqrt{r^3+\epsilon^2}}{\sqrt{2}\mu} \right|^{-1/n} + \mathcal{B} - \frac{1}{6}
\left[ \left( 1 - \frac{\epsilon^2}{r^3} \right) \pm
\frac{2\sqrt{2(r^3+\epsilon^2)}}{\mu} \frac{\epsilon^2}{r^3} \right] \right.
\nonumber\\
& \left. \hspace{0.8cm} - \frac{1-1/n}{6} \left[
\frac{\epsilon^{2/3}}{anc\mu^2\gamma} \frac{r^3-\epsilon^2}{2r^3} - \left( 1 -
\frac{\epsilon^2}{r^3} \right) \frac{\sqrt{2(r^3+\epsilon^2)}}{\mu} \left(
\frac{\sqrt{r^3+\epsilon^2}}{\sqrt{2}\mu} \mp 1 \right) \right] \right\}
\\
\label{Y}
Y = & -\frac{\mathcal{A}}{n} \frac{r^3+\epsilon^2}{\mu^2} \left| 1 \mp
\frac{\sqrt{r^3+\epsilon^2}}{\sqrt{2}\mu} \right|^{-2(1-1/n)}
\nonumber\\
& \times \left\{ 1 - \frac{e^{a\sigma}}{2}\frac{|W_0|}{|A_0|} \left| 1 \mp
\frac{\sqrt{r^3+\epsilon^2}}{\sqrt{2}\mu} \right|^{-1/n} + \mathcal{B} - \frac{1}{6}
\left[ \frac{\epsilon^{2/3}}{ancr^3\gamma} + 2 \left(1 + \frac{\epsilon^2}{r^3}
\right) \right] \right.
\nonumber\\
& \left. \hspace{0.8cm} - \frac{1-1/n}{6} \left[
\frac{\epsilon^{2/3}}{anc\mu^2\gamma} \frac{r^3-\epsilon^2}{2r^3} - \left( 1 -
\frac{\epsilon^2}{r^3} \right) \frac{\sqrt{2(r^3+\epsilon^2)}}{\mu} \left(
\frac{\sqrt{r^3+\epsilon^2}}{\sqrt{2}\mu} \mp 1 \right) \right] \right\} \, .
\end{align}
Note that we can write $Y$ using $X$ as
\begin{equation}
Y =  \mp \frac{\sqrt{r^3+ \epsilon^2}}{\sqrt{2}\mu} X
+ \frac{\mathcal{A}}{6n\mu^2} (r^3 + \epsilon^2) \left| 1 \mp
\frac{\sqrt{r^3+\epsilon^2}}{\sqrt{2}\mu} \right|^{-2(1-1/n)} \left\{ 1 +
\frac{\epsilon^{2/3}}{ancr^3\gamma} + \left[ 3 \mp
\frac{2\sqrt{2(r^3+\epsilon^2)}}{\mu} \right] \frac{\epsilon^2}{r^3} \right\}
\, .
\end{equation}

To estimate the stability near the tip, let us take the limit $r^3 \to \epsilon^2$,
i.e. very close to the end of the inflationary epoch. Then, from (\ref{X}) and
(\ref{Y}), we can see that
\begin{align}\label{Xlimit}
X \to & \pm \frac{2\mathcal{A}}{n}\frac{\epsilon}{\mu} \left| 1 \mp
\frac{\epsilon}{\mu} \right|^{-2(1-1/n)} \left[ 1 -
\frac{e^{a\sigma}}{2}\frac{|W_0|}{|A_0|} \left| 1 \mp \frac{\epsilon}{\mu}
\right|^{-1/n} + \frac{a}{6} (2\sigma - \gamma k_0) \mp \frac{2\epsilon}{3\mu}
\right] \, ,
\\
Y \to & -\frac{2\mathcal{A}}{n} \left( \frac{\epsilon}{\mu} \right)^2 \left| 1 \mp
\frac{\varepsilon}{\mu} \right|^{-2(1-1/n)} \left[ \frac{1}{3} -
\frac{e^{a\sigma}}{2}\frac{|W_0|}{|A_0|} \left| 1 \mp \frac{\epsilon}{\mu}
\right|^{-1/n} + \frac{a}{6}(2\sigma - \gamma k_0) -
\frac{1}{6anc\epsilon^{4/3}\gamma} \right]
\nonumber\\
& = \mp \frac{\epsilon}{\mu}X + \frac{\mathcal{A}}{3n} \left(
\frac{\epsilon}{\mu} \right)^2 \left| 1 \mp \frac{\epsilon}{\mu}
\right|^{-2(1-1/n)} \left[ 4 \left( 1 \mp \frac{\epsilon}{\mu} \right) +
\frac{1}{anc\epsilon^{4/3}\gamma} \right] \, .
\end{align}
Further, in this limit, all the eigenvalues in (\ref{massmatrix}) become $X+Y$, so
that for the angular stability along $\alpha_2$, $\alpha_3$ and $\alpha_4$ we
require that
\begin{equation}\label{stability}
X+Y = \left(1 \mp \frac{\epsilon}{\mu} \right) X + \frac{\mathcal{A}}{3n} \left(
\frac{\epsilon}{\mu} \right)^2 \left| 1 \mp \frac{\epsilon}{\mu}
\right|^{-2(1-1/n)} \left[ 4 \left( 1 \mp \frac{\epsilon}{\mu} \right) +
\frac{1}{anc\epsilon^{4/3}\gamma} \right] > 0 \, .
\end{equation}
To complete the analysis we therefore need extra information, e.g. the value of the
stabilized volume modulus at the tip $\sigma_0$ and the ratio $\epsilon/\mu$. Since
$\epsilon/\mu$ is the ratio of the size of the tip $\epsilon$ to the distance of the
stack of $D7$ branes to the tip $\mu$, one can easily tune it such that
$\epsilon/\mu < 1$. Therefore it is sufficient to check the positivity of
(\ref{Xlimit}). For this, we apply the results we establish in the following
appendix (\ref{VFsigmaF}) and (\ref{Defomega0}) related to $\sigma_0$: then
 the
terms in the square brackets of (\ref{Xlimit}) can be written as
\begin{equation}
1 - \frac{e^{a\sigma}}{2}\frac{|W_0|}{|A_0|} \left| 1 \mp \frac{\epsilon}{\mu}
\right|^{-1/n} + \frac{a}{6} (2\sigma - \gamma k_0) \mp \frac{2\epsilon}{3\mu}
\approx \frac{1}{2} - \frac{bs}{6} \mp \frac{2\epsilon}{3\mu} \approx \frac{1}{2} -
\frac{bs}{6} \, .
\end{equation}
Thus the angular stability depends on the product $bs$: if $bs > 3$ we obtain
negative sign while $bs < 3$ it becomes positive. Since we know that the power $b$
for our uplifting potential is either 2 (for $V_{D3\overline{D3}}$) or 3 (for
$V_{D\mathrm{-term}}$) and that $1 < s \lesssim \mathcal{O}(3)$, the product $bs$
lies in the range
\begin{equation}
2 < bs \lesssim \mathcal{O}(9) \, .
\end{equation}
We can therefore see
 that the condition for angular stability $bs
> 3$ can be naturally satisfied. Thus we conclude that $z_1^{(0)} =
-\sqrt{(r^3+\varepsilon^2)/2}$ is the stable trajectory we have been searching for
very near the tip.

\section{Derivation of approximated stabilized volume}
\label{app_vol}
\setcounter{equation}{0}
\paragraph{}

Having derived the angular stable trajectory $z_1=-\sqrt{(r^3+\epsilon^2)/2}$, we
are left with a two-field potential $V_F(r,\sigma) = V_\mathrm{KKLT}(r,\sigma) +
\Delta V_F(r,\sigma)$. In this appendix we will derive an approximate expression of
the stabilized volume $\sigma_\star(r)$ in term of the radial coordinate $r$ that is
given by (\ref{Mainapproxvolume2}).

\subsection{Stabilized volume at the tip}
\paragraph{}

Without taking into account of the uplifting term and the additional contribution
from the $D3$ brane position, we have the usual anti de Sitter minimum of KKLT
compactification, the stabilized volume modulus $\sigma_F$ is defined to be
\begin{equation}\label{SUSYmin}
\left. \frac{\partial{V}_F}{\partial\sigma} (r, \sigma)
\right|_{r=\epsilon^{2/3}, \sigma = \sigma_F} = 0 \, .
\end{equation}
Explicitly $\sigma_F$ can be given by the transcendental equation
\begin{equation}\label{2ndtermsigmaF}
\frac{|W_0|}{|A_0|}e^{a\sigma_F}\cG^{1/n}
= 1 + \frac{aU_F}{3} \, ,
\end{equation}
where $\cG$ is defined by (\ref{curlyG}) and
\begin{equation}\label{UF}
U_F = 2\sigma_F - \gamma k_0 \, .
\end{equation}
Thus at the tip, the potential is given by
\begin{equation}\label{VFsigmaF}
V_F(r=\varepsilon^{2/3},\sigma = \sigma_F) =\left.V_{\rm KKLT}\right|_{r=\epsilon^{2/3}}=
-\frac{a^2\kappa^2|A_0|^2e^{-2a\sigma_F}}{3U_F} \cG^{-2/n} \, .
\end{equation}
Notice that $\Delta V_F$ vanishes at $r=\epsilon^{2/3}$.

Now consider including the effect of uplifting term $V_D(r,\sigma)$ which we assume
to take the form in (\ref{DefVDuplifting}). We expect such a term contributes a
small shift to the stabilized volume at the tip of the throat
$\sigma_0=\sigma_F+\delta\sigma$, which is formally defined as
\begin{equation}\label{DefOmega0}
\left.\frac{\partial{(V_F+V_D)}}{\partial\sigma}\right|_{r=\varepsilon^{2/3},\sigma=\sigma_0}
\approx \left. \frac{\partial^2V_F}{\partial\sigma^2} \right|_{\sigma_F}\delta\sigma
+ \left. \frac{\partial{V_D}}{\partial\sigma} \right|_{\sigma_0} = 0 \, ,
\end{equation}
where
\begin{align}\label{approxdVD}
\left. \frac{\partial V_D}{\partial\sigma} \right|_{\sigma_0} = & -\frac{2b
V_D}{2\sigma_0-\gamma k_0}
\approx  -\frac{b V_D}{\sigma_F} \left[ 1 - (b+1)
\left(\frac{\delta\sigma}{\sigma_F} + \frac{\gamma k_0}{2\sigma_F} \right)
\right]\,.
\end{align}
Solving (\ref{DefOmega0}) and (\ref{approxdVD}), we obtain
\begin{equation}\label{approxdeltasigma}
\frac{\delta\sigma}{\sigma_F} \approx \left[ 1 - (b+1) \frac{\gamma k_0}{2\sigma_F}
\right] \left[ (b+1) + \left. \frac{\sigma_F^2}{bV_D}\frac{\partial^2
V_F}{\partial\sigma^2} \right|_{\sigma_F} \right]^{-1} \, .
\end{equation}
We can also find that
\begin{align}\label{VF''sigmaF}
\left.\frac{\partial^2V_F}{\partial\sigma^2}\right|_{r=\varepsilon^{2/3},\sigma=\sigma_F}
= & \frac{2a\kappa^2|A_0|^2e^{-2a\sigma_F}}{U_F^2} \cG^{-2/n} \left(
\frac{a^3U_F}{3} + \frac{5}{3}a^2 - \frac{16a}{3U_F} \right)
\nonumber\\
\approx & 2a^2 \frac{a^2\kappa^2|A_0|^2e^{-2a\sigma_F}}{3U_F} \cG^{-2/n}
\nonumber\\
= & 2a^2|V_F(r=\epsilon^{2/3},\sigma = \sigma_F)| \, ,
\end{align}
where we have used the fact that typically $\sigma_F \gg 1$. The shift of the
stabilized volume can then be approximated as
\begin{equation}\label{Defomega0}
\delta\sigma \approx \frac{bs}{2a^2\sigma_F}\, ,
\end{equation}
where the parameter $s$ is the uplifting ratio given by (\ref{Defs}).

\subsection{Radial dependence}
\paragraph{}

Now let us also take into account the dependence of the stabilized volume on the
brane position, denoted as $\sigma_\star(r)$. Formally this amounts to solving the
equation
\begin{equation}\label{DVFVD}
\left. \frac{\partial [V_F(r,\sigma)+V_D(r,\sigma)]}{\partial \sigma}
\right|_{\sigma_\star(r)}=0 \, .
\end{equation}
From (\ref{VKKLT2}), (\ref{DeltaVF2}) and (\ref{DefVDuplifting}) and the previous
analysis, the volume modulus $\sigma$ appears in both the exponential and the
polynomial, (\ref{DVFVD}) is thus in fact a transcendental equation. To simplify the
analysis, one notices that in the large $\sigma_F/\sigma_0$ limit, one can
approximate $\sigma_\star(r)$ in the polynomial by $\sigma_0$ \cite{Baumann1}, as
the difference is exponentially suppressed. Then we are left with a quadratic
equation of $X \equiv \exp(-a\sigma_\star)$ given by
\begin{equation}\label{Volumeequation}
A_2X^2 + A_1X - A_0 = 0 \, ,
\end{equation}
so that
\begin{equation}\label{sigmastarsol}
\sigma_\star = \frac{1}{a} \log \left[ \frac{2A_2}{A_1} \left( -1 - \sqrt{1 +
\frac{4A_0A_2}{A_1^2}} \right)^{-1} \right] \, .
\end{equation}
Two comments are in order: first, we note that
\begin{equation}
A_0 = \frac{2bD(\sepy)}{U_{\sigma_0}^{b+1}} \propto \frac{1}{\sigma_0^{b+1}} \, ,
\end{equation}
with $b = 2$ or 3. Therefore, unless we care for corrections beyond
$\mathcal{O}(1/\sigma_0^{b+1})$, the factor $4A_0A_2/A_1^2$ in the square root does
not alter the leading contribution if there exist terms up to
$\mathcal{O}(1/\sigma_0^b)$, which is indeed the case as will be shown. Second,
since we are interested in the region very close to $r = \epsilon^{2/3}$, the
primary expansion parameter will be $r - \epsilon^{2/3}$ which we denote by
$x$ below. Thus we expect
\begin{align}
\frac{2A_2}{A_1} \left( -1 - \sqrt{1 + \frac{4A_0A_2}{A_1^2}} \right)^{-1} = &
-\frac{A_2}{A_1} + \mathcal{O} \left( \frac{1}{\sigma_0^{b+1}} \right)
\nonumber\\
= & e^{a\sigma_F} \left\{ \left[ \mathcal{O}(1) + \mathcal{O} \left(
\frac{1}{\sigma_0} \right) + \cdots + \mathcal{O} \left( \frac{1}{\sigma_0^{b}}
\right) \right] \right.
\nonumber\\
& \left. \hspace{1cm} + \left[ \mathcal{O}(1) + \mathcal{O} \left(
\frac{1}{\sigma_0} \right) + \cdots + \mathcal{O} \left( \frac{1}{\sigma_0^{b}}
\right) \right] x + \cdots \right\} \, .
\end{align}
After some calculations, we can find that schematically
\begin{equation}
-\frac{A_2}{A_1} = e^{a\sigma_F} \left( c_0 + c_1 x \right) \, ,
\end{equation}
where the coefficients are given such that
\begin{align}
c_0 = & 1 + \mathcal{O} \left( \frac{1}{\sigma_F^2} \right) \, ,
\\
c_1 = & \mp \frac{3\epsilon^{1/3}}{4n\mu} \left|1\pm\frac{\epsilon}{\mu}\right|^{-1}
+ \mathcal{O} \left( \frac{1}{\sigma_F} \right) \, ,
\end{align}
so that
\begin{equation}\label{Defsigmastar1}
\sigma_\star \approx \sigma_F \left( 1 + \frac{c_1}{a\sigma_F}x \right) \, .
\end{equation}
One notices that in contrast with Ref.~\cite{Baumann1}, where the leading radial
dependence enters at order $r^{3/2}$, here with the deformation parameter
$\epsilon\neq 0$, we only have a rational expansion. Furthermore as $\sigma_F$ and
$\sigma_0$ only differs at order $1/\sigma_F$, we can replace $\sigma_F$ by
$\sigma_0$ in (\ref{Defsigmastar1}).

\section{Calculations of the slow-roll parameter}
\label{app_slow-roll}
\setcounter{equation}{0}
\paragraph{}

In this section we present explicit calculation of the slow-roll parameter
$\varepsilon$ given in the main text. From (\ref{eq:KahlerPotential}),
(\ref{Mainapproxvolume2}), (\ref{C1}) and (\ref{asymptotic_k}), the derivatives of
$\sigma_\star(r)$ and $U[r,\sigma_\star(r)]$ with respect to the radial coordinate
$r$ are given by
\begin{align}
\frac{\partial\sigma_\star(r)}{\partial{r}} = & \frac{3\epsilon^{1/3}}{4an\mu}
\cG \, ,
\\
\frac{\partial U[r,\sigma_\star(r)]}{\partial{r}} = & \frac{3\epsilon^{1/3}}{2an\mu}
\cG - \frac{3c\gamma}{\epsilon^{2/3}}r^2 \, ,
\end{align}
respectively, where $\cG$ is again given by (\ref{curlyG}). Given these, now let us
calculate the derivatives of $\mathbb{V}$ with respect to $r$ from
(\ref{VKKLTsingle}), (\ref{DeltaVFsingle}) and (\ref{VDsingle}). For notational
simplicity, from below let us denote
\begin{equation}\label{Defgr}
g(r) \equiv  1 + \frac{\sqrt{r^3+\epsilon^2}}{\sqrt{2}\mu}\, .
\end{equation}
We can find, after some simple calculations, that the first derivatives with
respect to $r$ are given by
\begin{align}
\frac{\partial{V_\mathrm{KKLT}}}{\partial{r}} = & V_\mathrm{KKLT} \left\{
-\frac{3\epsilon^{1/3}}{n\mu} \mathcal{G} \left( \frac{1}{2} + \frac{1}{aU} \right)
+ \frac{6c\gamma}{\epsilon^{2/3}U}r^2 + \frac{3r^2}{2n\mu^2[g(r) - 1]g(r)} \right\}
\nonumber\\
& + \frac{2\kappa^2a|A_0|^2e^{-2a\sigma_\star}}{U^2} \left[ g(r) \right]^{2/n}
\nonumber\\
& \hspace{0.5cm} \times \left\{ \frac{|W_0|}{|A_0|}e^{a\sigma_\star} \left[ g(r)
\right]^{-1/n} \left[ -\frac{3\epsilon^{1/3}}{4n\mu} \mathcal{G} +
\frac{3r^2}{4n\mu^2[g(r)-1]g(r)} \right] + \frac{a}{2} \left[
\frac{\epsilon^{1/3}}{2an\mu} \mathcal{G} + c\gamma \frac{\epsilon^{10/3}}{r^4}
\right] \right\} \, ,
\label{dVKKLTdr}\\
\frac{\partial\Delta{V_F}}{\partial{r}} = & \Delta{V_F} \left\{
-\frac{3\epsilon^{1/3}}{n\mu} \mathcal{G} \left( \frac{1}{2} + \frac{1}{aU} \right)
+ \frac{6c\gamma}{\epsilon^{2/3}U}r^2 + \left( \frac{1}{n} - 1 \right)
\frac{3r^2}{2\mu^2[g(r)-1]g(r)} \right\}
\nonumber\\
& + \frac{\kappa^2|A_0|^2e^{-2a\sigma_\star}}{n^2U^2} \left[ g(r) \right]^{2(1/n-1)}
\nonumber\\
& \hspace{0.5cm} \times \left\{ \left( 1 - \frac{\epsilon^2}{r^3} \right)
\frac{(an-1)g(r)+1}{g(r)-1} \frac{r^2}{2\mu^2} + \frac{\epsilon^2}{r^4} \left[
\frac{\epsilon^{2/3}}{2c\mu^2\gamma} - 2an [g(r)-1]g(r) \right] \right\} \, ,
\label{dDVFdr}\\
\frac{\partial{V_D}}{\partial{r}} = & \frac{1}{U^b} \left\{
\frac{\partial{D(\sepy)}}{\partial{r}} -\frac{bD(\sepy)}{U} \left[
\frac{3\epsilon^{1/3}}{2an\mu} \mathcal{G} - \frac{3c\gamma}{\epsilon^{2/3}}r^2
\right] \right\} \, .
\label{dVDdr}
\end{align}
As can be read from the above expressions, in general $\varepsilon$ 
is a complex functions of $r$. To catch a clearer feeling, it would be useful to
evaluate them at the tip, i.e. at $r = \epsilon^{2/3}$. (\ref{dVKKLTdr}),
(\ref{dDVFdr}) and (\ref{dVDdr}) then become
\begin{align}
\left. \frac{\partial{V_\mathrm{KKLT}}}{\partial{r}} \right|_{r = \epsilon^{2/3}} =
& \left| V_\mathrm{KKLT}|_{r = \epsilon^{2/3}} \right| \frac{3}{U_F} \left[
\frac{3\epsilon^{1/3}}{2an\mu}\cG -
c\gamma\epsilon^{2/3} \right] \, ,
\\
\left. \frac{\partial\Delta{V_F}}{\partial{r}} \right|_{r = \epsilon^{2/3}} = &
\left| V_\mathrm{KKLT}|_{r = \epsilon^{2/3}} \right| \frac{3}{a^2n^2U_F} \cG^2
\left[ \frac{\epsilon^{2/3}}{2c\mu^2\gamma} - 2an \frac{\epsilon}{\mu} \cG^{-1}
\right] \epsilon^{-2/3} \, ,
\\
\left. \frac{\partial{V_D}}{\partial{r}} \right|_{r = \epsilon^{2/3}} = &
\frac{1}{U_F^b} \left\{ \left. \frac{\partial{D}}{\partial{r}}
\right|_{r=\epsilon^{2/3}} - \frac{bD|_{r=\epsilon^{2/3}}}{U_F} \left[
\frac{3\epsilon^{1/3}}{2an\mu} \cG -
3c\gamma\epsilon^{2/3} \right] \right\} \, ,
\end{align}
where $V_\mathrm{KKLT}|_{r = \epsilon^{2/3}}$ is given by (\ref{VFsigmaF}) and $U_F$
by (\ref{UF}). Hence, we have
\begin{align}\label{Vrderiv}
\left. \frac{\partial{{\mathbb{V}}}}{\partial{r}} \right|_{r = \epsilon^{2/3}} = &
\frac{3\left| V_\mathrm{KKLT}|_{r = \epsilon^{2/3}} \right|}{U_F} \left\{ \left[
\frac{3\epsilon^{1/3}}{2an\mu} \cG -
c\gamma\epsilon^{2/3} \right] \right.
\left. + \frac{\cG^2}{a^2n^2\epsilon^{2/3}} \left[
\frac{\epsilon^{2/3}}{2c\mu^2\gamma} - 2an \frac{\epsilon}{\mu} \cG^{-1} \right] \right\}
\nonumber\\
& + \frac{1}{U_F^b} \left\{ \left. \frac{\partial{D}}{\partial{r}}
\right|_{r=\epsilon^{2/3}} - \frac{bD|_{r=\epsilon^{2/3}}}{U_F} \left[
\frac{3\epsilon^{1/3}}{2an\mu} \cG -
3c\gamma\epsilon^{2/3} \right] \right\} \, .
\end{align}
Further, from (\ref{Defs}) we can relate $V_D|_{r=\epsilon^{2/3}}$ with
$V_\mathrm{KKLT}|_{r=\epsilon^{2/3}}$ by
\begin{equation}
V_D|_{r=\epsilon^{2/3}} = \frac{D|_{r=\epsilon^{2/3}}}{U_F^b} =
s|V_\mathrm{KKLT}|_{r=\epsilon^{2/3}}| \, .
\end{equation}
This further simplifies (\ref{Vrderiv}) into
\begin{align}
\left. \frac{\partial{{\mathbb {V}}}}{\partial{r}} \right|_{r = \epsilon^{2/3}} = &
|V_\mathrm{KKLT}|_{r=\epsilon^{2/3}}| \left\{ \frac{3-sb}{U_F} \left[
\frac{3\epsilon^{1/3}}{4\pi\mu} \cG - c\gamma\epsilon^{2/3} \right] +
\frac{2sbc\gamma}{U_F}\epsilon^{2/3} + \frac{s}{D} \left.
\frac{\partial{D}}{\partial{r}} \right|_{r=\epsilon^{2/3}} \right.
\nonumber\\
& \hspace{2.5cm} \left. + \frac{3\cG^2}{4\pi^2U_F} \left[
\frac{\epsilon^{2/3}}{2c\mu^2\gamma} - 4\pi \frac{\epsilon}{\mu} \cG^{-1} \right]
\epsilon^{2/3} \right\} \, .
\end{align}
Likewise, using the fact that $\Delta{V_F}|_{r = \epsilon^{2/3}} = 0$,
\begin{align}\label{dlogVdr}
{\mathbb{V}}|_{r = \epsilon^{2/3}} = & V_\mathrm{KKLT}|_{r = \epsilon^{2/3}} +
\Delta{V_F}|_{r = \epsilon^{2/3}} + V_D|_{r = \epsilon^{2/3}}
\nonumber\\
= & V_\mathrm{KKLT}|_{r = \epsilon^{2/3}} + V_D|_{r = \epsilon^{2/3}}
\nonumber\\
= & (s-1) |V_\mathrm{KKLT}|_{r = \epsilon^{2/3}}| \, .
\end{align}
Thus, at the tip we obtain
\begin{align}\label{VrV}
\left. \frac{\partial{\mbV}/\partial{r}}{\mbV} \right|_{r = \epsilon^{2/3}} = &
\frac{1}{s-1} \left\{ \frac{3-sb}{U_F} \left[ \frac{3\epsilon^{1/3}}{4\pi\mu}
\cG - c\gamma\epsilon^{2/3} \right] +
\frac{2sbc\gamma}{U_F}\epsilon^{2/3} + \frac{s}{D} \left.
\frac{\partial{D}}{\partial{r}} \right|_{r=\epsilon^{2/3}} \right.
\nonumber\\
& \hspace{1.2cm} \left. + \frac{3\cG^2}{4\pi^2U_F} \left[
\frac{\epsilon^{2/3}}{2c\mu^2\gamma} - 4\pi \frac{\epsilon}{\mu} \cG^{-1} \right]
\epsilon^{2/3} \right\} \, .
\end{align}
Note that $D$ contains the factor $\sepy$ and that its derivative is given by
\begin{equation}
\frac{\partial\sepy}{\partial{r}} = 2(r - \epsilon^{2/3}) \, ,\label{Dsepydr}
\end{equation}
we can easily see that the term involving $\partial{D(\sepy)}/\partial{r}$ in fact
vanishes at $r=\epsilon^{2/3}$ for $V_{D3\overline{D3}}$ with $b=2$, whereas for
$V_{D\mathrm{-term}}$ with $b=3$ such term vanishes identically as $D(\sepy)=v_D$.
In other words, $\varepsilon_\tip$ for the two different uplifting mechanisms only
differ in $b$. The canonically normalized inflaton near the tip is identified as in
(\ref{Inflatontau}) and $\partial_\phit = (\partial r/\partial\phit)\partial_r$.
Note that $\epsilon$ and $\mu$, $\gamma$, and $r$ have mass dimension of --3/2, 2
and --1 respectively, so it can be easily seen that
$(\partial{\mbV}/\partial{r})/\mbV$ has mass dimension 1 by counting the
dimensionful parameters.

Since (\ref{VrV}) is clearly finite, from the chain rule (\ref{dphitdrs}) one can
easily see that the slow-roll parameter $\varepsilon$ in fact vanishes identically
at the tip $r=\epsilon^{2/3}$, or $\phit=0$. However we can expand around the tip
and obtain the 
lowest order approximation as
\begin{equation}\label{epsilontip}
\varepsilon(r) \approx \left. \frac{\mpl^2}{3T_3\epsilon^{4/3}} \left(
\frac{\partial{\mbV}/\partial{r}}{\mbV} \right)^2 \right|_{r = \epsilon^{2/3}}
\left( r^2-\epsilon^{4/3}\right) \, ,
\end{equation}
with $(\partial{\mbV}/\partial{r})/\mbV|_{r=\epsilon^{2/3}}$ given by (\ref{VrV}).
We can express $\varepsilon(r)$ in terms of the compactification parameters
describing the bulk and throat geometries \cite{Baumann1}. Several useful
expressions are
\begin{align}
\label{tipwarpfactor}
\frac{\hat{r}_\mathrm{UV}}{\epsilon^{2/3}} = &
\sqrt{\frac{3}{2}}\frac{r_\mathrm{UV}}{\epsilon^{2/3}} = \exp \left( \frac{2\pi
K}{3g_sM} \right) \equiv a_0^{-1} \, ,
\\
Q_\mu \equiv & \frac{r_\mathrm{UV}}{r_\mu} = \frac{r_\mathrm{UV}}{(2\mu^2)^{1/3}} =
\frac{2^{1/6}}{3^{1/2}a_0} \left( \frac{\epsilon}{\mu} \right)^{2/3} \, ,
\\
\gamma \approx & \frac{B_4}{B_6} \frac{2\log{Q_\mu}}{3\pi r_\mathrm{UV}^2} =
\frac{B_4}{B_6} \frac{2\log{Q_\mu}}{3\pi(2\mu^2)^{2/3}Q_\mu^2} \, ,
\\
\mpl^2 = & \frac{T_3^2}{\pi}V_6^w = \frac{3}{8}NB_6T_3 (2\mu^2)^{2/3}Q_\mu^2 \, ,
\\
U_F \approx & 2\sigma_F \approx \frac{3N}{2\pi}B_4\log{Q_\mu} \, .
\end{align}
Here $\hat{r}_\mathrm{UV}$ denotes the ultraviolet cutoff radius where the deformed
conifold is attached to the bulk Calabi-Yau, $r_\mu$ is the minimal radius of $D7$,
and $B_4$ and $B_6$ denote the contributions of the throat to the warped volume of
the wrapped four cycle and to the total warped volume of the compact space $V_6^w$,
respectively. Substituting these into (\ref{VrV}) and (\ref{epsilontip}) and after a
little calculations, we find
\begin{align}\label{dlogvdrtip}
\left. \frac{\partial{\mbV}/\partial{r}}{\mbV} \right|_{r = \epsilon^{2/3}} = &
\frac{\mu^{-2/3}}{s-1} \left\{ \frac{3-sb}{3NB_4\log{Q_\mu}} \left[ \frac{3}{2}
3^{1/12} \left( \frac{a_0Q_\mu}{c} \right)^{1/2} \left(1 + 3^{1/4} \left(
\frac{a_0Q_\mu}{c} \right)^{3/2} \right) \right.\right.
\nonumber\\
& \left. \hspace{4cm} - \frac{B_4}{B_6} \frac{2\cdot2^{1/3}c\log{Q_\mu}}{3Q_\mu^2}
3^{1/12}\left( \frac{a_0Q_\mu}{c} \right) \right]
+ \frac{4\cdot2^{1/3}sbc}{9NB_6Q_\mu^2} 3^{1/6} \left(
\frac{a_0Q_\mu}{c} \right)
\nonumber\\
& \hspace{1.2cm} + \frac{2}{NB_4\log{Q_\mu}}\left[ 1 + 3^{1/4}\left(
\frac{a_0Q_\mu}{c} \right)^{3/2} \right]^{-2}
\nonumber\\
& \left. \hspace{1.5cm} \times \left[ \frac{B_6}{B_4}
\frac{3Q_\mu^2}{8\cdot2^{1/3}c\log{Q_\mu}} - 3^{1/12} \left( \frac{a_0Q_\mu}{c}
\right)^{1/2} \left( 1 + 3^{1/4}\left( \frac{a_0Q_\mu}{c} \right)^{3/2} \right)
\right] \right\} \, .
\end{align}
We can also observe from above that for the two different uplifting mechanisms, the
value of $\varepsilon(r)$ only differ in $b$.

\end{document}